%% file: main.ARXIV.tex
\documentclass[%
 aip,
 jcp,
 amsmath,amssymb,
reprint,
twocolumn
]{revtex4-1}

\usepackage{float}
\usepackage[hypertexnames=false]{hyperref}
\hypersetup{
    colorlinks = true,
    allcolors = {blue}
}
\usepackage[all]{hypcap}   
\usepackage{enumitem}
\usepackage{graphicx}
\usepackage{wrapfig}
\begin{document}

\title{The Many Faces of Heterogeneous Ice Nucleation: Interplay Between Surface Morphology and Hydrophobicity}

\author{Martin Fitzner}
\affiliation{London Centre for Nanotechnology, Department of Chemistry and Thomas Young Centre, University College London, 20 Gordon Street, London WC1H 0AJ, United Kingdom}

\author{Gabriele C. Sosso}
\affiliation{London Centre for Nanotechnology, Department of Chemistry and Thomas Young Centre, University College London, 20 Gordon Street, London WC1H 0AJ, United Kingdom}

\author{Stephen J. Cox}
\affiliation{London Centre for Nanotechnology, Department of Chemistry and Thomas Young Centre, University College London, 20 Gordon Street, London WC1H 0AJ, United Kingdom}

\author{Angelos Michaelides}
\affiliation{London Centre for Nanotechnology, Department of Chemistry and Thomas Young Centre, University College London, 20 Gordon Street, London WC1H 0AJ, United Kingdom}
\email{angelos.michaelides@ucl.ac.uk}
\date{15 October 2015}

\begin{abstract}
\begin{wrapfigure}{r}{0.4\textwidth}
	\hspace{-0.19\linewidth}
  \includegraphics[width=1.0\linewidth]{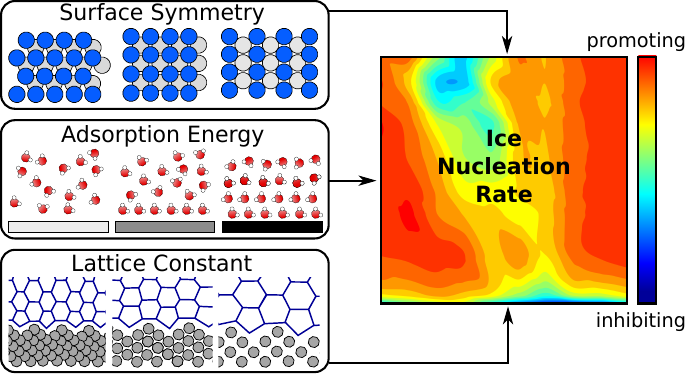}%
\end{wrapfigure}
What makes a material a good ice nucleating agent? Despite the importance of heterogeneous ice nucleation to a variety of fields, from cloud science to microbiology, major gaps in our understanding of this ubiquitous process still prevent us from answering this question. In this work, we have examined the ability of generic crystalline substrates to promote ice nucleation as a function of the hydrophobicity and the morphology of the surface. Nucleation rates have been obtained by brute-force molecular dynamics simulations of coarse-grained water on top of different surfaces of a model fcc crystal, varying the water-surface interaction and the surface lattice parameter. It turns out that the lattice mismatch of the surface with respect to ice, customarily regarded as the most important requirement for a good ice nucleating agent, is at most desirable but not a requirement. On the other hand, the balance between the morphology of the surface and its hydrophobicity can significantly alter the ice nucleation rate and can also lead to the formation of up to three different faces of ice on the same substrate. We have pinpointed three circumstances where heterogeneous ice nucleation can be promoted by the crystalline surface: i) the formation of a water overlayer that acts as an in-plane template; ii) the emergence of a contact layer buckled in an ice-like manner; and iii) nucleation on compact surfaces with very high interaction strength. We hope that this extensive systematic study will foster future experimental work aimed at testing the physiochemical understanding presented herein.
\end{abstract}

\maketitle

\section{Introduction}
The formation of ice influences our everyday experience as well as a variety of scenarios, ranging from global phenomena like climate change~\cite{murray_ice_2012,bartels-rausch_chemistry:_2013} to processes happening at the nanoscale, like intracellular freezing~\cite{mazur_cryobiology:_1970,lintunen_anatomical_2013}. It is surprisingly difficult to observe ice crystallization from pure supercooled water, because the pure liquid can be cooled to $-40^\circ$C without freezing~\cite{hegg_nucleation_2009,murray_ice_2012}. In fact, ice nucleation in nature happens mostly thanks to the presence of foreign particles~\cite{bartels-rausch_ice_2012}, ranging from biological compounds to crystalline surfaces~\cite{murray_ice_2012}. Such spectacular diversity calls for an obvious question: what is it that makes a material a good ice nucleating agent (INA)? The vast body of experimental and theoretical work undertaken within the last few decades in order to answer this seemingly trivial query proves that our understanding of heterogeneous ice nucleation is far from satisfactory. 
 
Recently, a number of excellent experimental works have succeeded in determining which materials can effectively promote heterogeneous ice nucleation, mostly by measuring ice nucleation temperatures or rates, see e.g.\citenum{finnegan_new_2003,michaelides_ice_2007,zobrist_heterogeneous_2008,zimmermann_ice_2008,murray_heterogeneous_2010,ehre_water_2010,murray_heterogeneous_2011,knopf_stimulation_2011,tobo_impacts_2012,li_investigating_2012,fu_ice_2015,whale_ice_2015}. By doing so, the ice nucleating abilities of a large variety of materials has been characterized~\cite{pruppacher1997microphysics,murray_ice_2012}. This knowledge can for instance be used to decipher and explain the different contributions to ice nucleation in the atmosphere~\cite{pratt_situ_2009,pratt_situ_2010,hoose_heterogeneous_2012,atkinson_importance_2013,hiranuma_ice_2015}. However, experiments currently do not provide information into the molecular details of individual ice nucleation events. Because of the length scale involved (nm), insights into the nucleation process can be obtained instead from computer simulations. 
And indeed, in the last few years a handful  of computational studies have been successful in simulating heterogeneous ice nucleation~\cite{moore_freezing_2010,yan_molecular_2012,hudait_ice_2014,zielke_molecular_2014,zhang_impact_2014,singh_characterization_2014,lupi_does_2014,lupi_heterogeneous_2014,reinhardt_effects_2014,fraux_note:_2014,cabriolu_ice_2015,cox_peeling_2015,cox_controlling_2015}. This indicates that the time is now ripe for furthering our understanding of the microscopic factors that make a material a good INA. Nevertheless, even being able to explore heterogeneous ice nucleation with simulation approaches may not be enough to understand \textit{a priori} whether and why a material will be a good INA or not. This is because many different ingredients like the morphology of the surface~\cite{hu_ice_2007,cox_non-hexagonal_2012,pirzadeh_molecular_2013,cox_microscopic_2014}, its hydrophobicity~\cite{michaelides_ice_2007,lupi_does_2014,cox_controlling_2015,Factorovich_hydrogen_2015}, local electric fields~\cite{hribar_how_2002,reveles_h2o_2007,ehre_water_2010,yan_heterogeneous_2011,yan_molecular_2012}, preferential nucleation sites or surface roughness~\cite{lupi_heterogeneous_2014,singh_characterization_2014,nistor_interface-limited_2014}, can simultaneously impact on both the molecular mechanism and the resulting nucleation rate.

The two most discussed ``requirements'' for an effective INA are perhaps the crystallographic match with respect to bulk ice and the strength of the water-surface interaction. The former was introduced by Turnbull and Vonnegut~\cite{turnbull_nucleation_1952} in order to characterize the catalytic potential of a surface regarding heterogeneous nucleation. If the atomic arrangements in the contact region are similar, a disregistry or lattice mismatch $\delta$ between the corresponding surface unit cells can be defined in a simplified manner as:
\begin{equation}
\label{eqn.mismatch}
	\delta = \frac{a_\mathrm{s} - a_\mathrm{i}}{a_\mathrm{i}}
\end{equation}
where $a_\mathrm{s}$ and $a_\mathrm{i}$ are the lattice parameters of the surface unit cells of the substrate and a certain face of ice. The idea of a small lattice mismatch $\delta$ being at the heart of the INA efficacy dates back to the forties, when the ice nucleating capabilities of AgI, featuring only $\delta \approx 2$~\% for the basal face, came to light~\cite{vonnegut_nucleation_1947,pruppacher1997microphysics}. Even though both experiments ~\cite{zettlemoyer_surface_1961,finnegan_new_2003} and recent simulations~\cite{reinhardt_effects_2014,fraux_note:_2014} have seriously challenged the validity of this concept and most importantly its generality with respect to other materials~\cite{cox_non-hexagonal_2012,cox_microscopic_2014}, a small lattice mismatch is still considered as the primary attribute of an efficient INA. In the case of bacterial ice nucleating proteins ice-matching patterns have even been used as an \textit{a priori} assumption to infer the three dimensional structure of the residues from the DNA sequence describing the protein ~\cite{kajava_model_1993,murakami_molecular_2012}.

Concerning the water-surface interaction, or the hydrophobicity/hydrophilicity of a surface, in the last two decades a number of experimental studies investigating ice formation on soot~\cite{gorbunov_ice_2001,dymarska_deposition_2006,karcher_insights_2007,pratt_situ_2010,friedman_ice_2011,murray_ice_2012,bingemer_atmospheric_2012} have prompted a debate about whether a correlation exists between the hydrophilicity of carbonaceous surfaces and their efficacy as INAs. This is a challenging issue, because in most cases the role of the hydrophobicity cannot be disentangled from the influence of the lattice mismatch and surface morphology. As an example, the oxidation of soot taking place in atmospheric aerosols modifies both the hydrophilicity and the morphology of the particles at the nanoscale~\cite{cabrera-sanfelix_dissociative_2007}. Furthermore, Lupi and Molinero~\cite{lupi_does_2014} found that an increase in hydrophilicity showed adverse effects when it was accomplished by adding OH groups as opposed to just increasing the water-surface interaction strength. And indeed, recent experiments by Whale \textit{et al.}~\cite{whale_ice_2015} provide some tentative support for this hypothesis. Cox \textit{et al.} recently investigated the dependence of the ice nucleation rate as a function of hydrophilicity in the case of model nano-particles~\cite{cox_peeling_2015}. They found a similar interaction range for both a fcc and a graphene-like particle where nucleation is enhanced, leading to a rule-of-thumb for an optimal adsorption strength. They also showed~\cite{cox_controlling_2015} how a simple modification of the surface morphology could lead to a significant change of nucleation rates, demonstrating the potential of atomic-scale control of nucleation.

As far as we know, the interplay between the hydrophobicity and morphology of the surface has not been systematically studied at the molecular scale. In this work, we fill this shortfall by investigating ice formation on top of a generic fcc crystal as a function of both the strength of the water-surface interaction and the morphology, taking into account the (111), (100), (110) and (211) surfaces. 
Strikingly different nucleation scenarios emerge according to the balance between the morphology of the surface and its hydrophobicity, thus demonstrating that the lattice mismatch alone cannot be deemed as the key player in promoting nucleation on crystalline surfaces. In addition, we have found that up to three different faces of ice can nucleate on top of the same surface, and that the microscopic motivation at the heart of the heterogeneous nucleation process is not unique, but actually changes according to both the water-surface interaction and the morphology of the surface. We propose three microscopic factors that lead to enhancement of the nucleation rates: i) the formation of a water overlayer that acts as an in-plane template; ii) the emergence of a contact layer buckled in an ice-like manner; and iii) enhanced nucleation on compact surfaces with very high adsorption energy. 

The remainder of this paper is organized as follows: Section \ref{sec.methods} describes the computational setup (\ref{sec.methods.1}) together with how we obtained nucleation rates and an assessment of finite-size effects (\ref{sec.methods.2}). In section \ref{sec.results} we present the nucleation rates for all the different surfaces as a function of adsorption energy and lattice constant. From this data we shall extract and discuss the general trends that emerge (\ref{sec.results.1}). The following subsection \ref{sec.results.2} presents the three different scenarios we propose as driving forces behind the nucleation promotion. We then discuss further insight and future perspectives for improved heterogeneous ice nucleation simulations and experiments that could test the suggestions made here in subsection \ref{sec.results.3}. Finally, the key results and observations are summarized in section \ref{sec.conclusions}.

\section{Methods}
\label{sec.methods}

\subsection{System and Computational Methods}
\label{sec.methods.1}
\begin{figure}[t]
\centerline{\includegraphics[width=8.4cm]{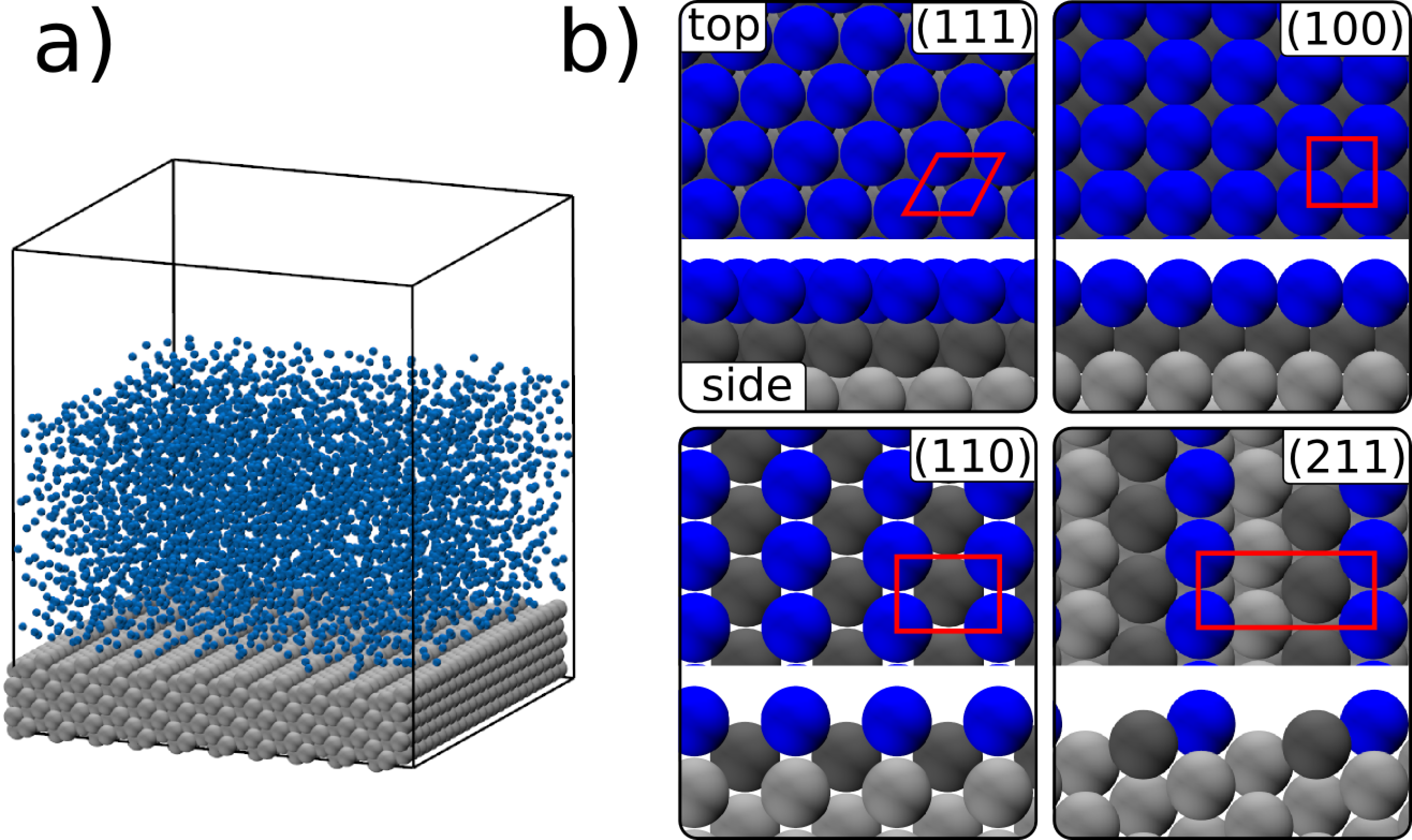}} 
\caption{a) Example of a simulation box used in a heterogeneous ice nucleation run. The coarse-grained water molecules are depicted as blue spheres while surface atoms are gray. The average box dimensions were $60\times60\times70$~\r{A}. b) Top and side view of the four crystalline surfaces considered. Atoms are colored according to their z-coordinate. Red boxes highlight the symmetry of the surface unit cells.} 
\label{FIG_1_SURFACE_LOOK} 
\end{figure}

We considered slab models of crystalline surfaces covered in a water film (see Figure~\ref{FIG_1_SURFACE_LOOK}a) including 4000 water molecules represented by the coarse-grained mW model~\cite{molinero_water_2009}. This specific water model has excellent structural properties and a melting point close to experiment~\cite{molinero_water_2009} but since it is monoatomic it exhibits faster dynamics which in turn allows for brute-force simulations of nucleation~\cite{moore_ice_2010,moore_is_2011,johnston_crystallization_2012,lupi_does_2014,lupi_heterogeneous_2014,cox_peeling_2015,cox_controlling_2015}. The water film is $\sim 35$~\r{A} thick, which is enough so that the density is converged to the bulk homogeneous value at $\sim 12$~\r{A} above the interface. 

We note that in general in this study we do not aim to mimic a specific system but to extract instead generic insight and trends from idealized model substrates. To this end we have taken into account four different crystallographic planes of a generic fcc crystal, namely the (111), (100), (110) and the (211) surfaces, which exhibit significant differences in terms of atomic roughness and the symmetry of the outer crystalline layer (see Figure~\ref{FIG_1_SURFACE_LOOK}b). For each of the above mentioned surfaces, we have built a dataset of ten different slabs varying the fcc lattice parameter $a_\mathrm{fcc}$ from 3.52 to 4.66~\r{A}~\cite{note_built}. This range encompasses the majority of fcc metals. The interaction of the water with the substrate is given by a truncated Lennard-Jones potential:
\begin{equation}
	U(r) =  \begin{cases}
				4\epsilon \left[ \Big(\frac{\sigma}{r}\Big)^{12} - \Big(\frac{\sigma}{r}\Big)^{6} \right] & r < r_c\\
                0 & r \geq r_\mathrm{c}
			\end{cases}
\end{equation}
where $r$ is the distance between a water oxygen and a surface atom. The cutoff distance was set to $r_\mathrm{c} = 7.53$~\r{A}. 

To measure the interaction strength of water with the surface the adsorption energy $E_\mathrm{ads}$ of a single water molecule was computed. In order to vary this quantity $\epsilon$ and $\sigma$ were changed accordingly.
$E_\mathrm{ads}$ was computed by minimizing the potential energy of a single water molecule on top of the surface. In this manner well defined adsorption energies can be determined for the (111), (100) and (110) surfaces since only one adsorption site is found by the minimization algorithm. However, for the (211) geometry multiple adsorption sites with considerable energy differences were found~\cite{note_min}. For this reason we have chosen to assign every $(a_\mathrm{fcc},E_\mathrm{ads})$ combination for the (211) orientation the same $(\epsilon,\sigma)$ pair as for the (111) surface. This is also motivated by the (111) terrace exhibited by the (211) surface. The final adsorption energy for the (211) geometry as reported in Figure~\ref{FIG_5_NUCLEATION_MAPS} is the arithmetic average of the different adsorption energies found on this particular surface. The averaged results deviate by ca. 5~\% from those for the (111) surface, e.g. the highest $E_\mathrm{ads}$ on (111) is around $12.76$~kcal/mol while the average value for the (211) surface with the same $(\epsilon,\sigma)$ parameters is $13.18$~kcal/mol. 

\subsection{Obtaining Nucleation Rates}
\label{sec.methods.2}
\begin{figure}[t] 
\centerline{\includegraphics[width=8.5cm]{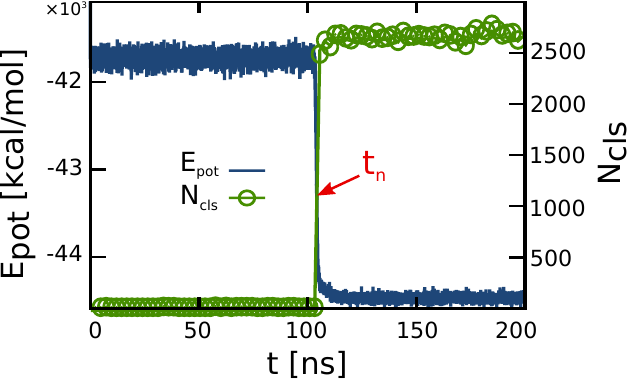}}
\caption{An illustration of how the nucleation induction time $t_\mathrm{n}$ is established by monitoring the change in the potential energy $E_\mathrm{pot}$ in blue. The green data shows the number of water molecules within the biggest ice-like cluster $N_\mathrm{cls}$~\cite{note_orderparameter} and that the jump in $N_\mathrm{cls}$ coincides with nucleation. The data refers to the (111) surface for $E_\mathrm{ads}$=1.04~kcal/mol and $a_\mathrm{fcc}=4.16$~\r{A}.}
\label{FIG_2_EPOT_VS_NCLS} 
\end{figure}

Heterogeneous ice nucleation events have been simulated by means of brute-force molecular dynamics (MD) simulations, employing the LAMMPS simulation package~\cite{plimpton_fast_1995}. We follow a similar protocol to the one of Cox \textit{et al.}~\cite{cox_controlling_2015}. A time step of 10~fs has been used with periodic boundary conditions in the xy-plane while sampling the NVT canonical ensemble with a chain of 10 Nos{\'e}-Hoover thermostats~\cite{nose_unified_1984,martyna_nosehoover_1992} with a relaxation time of 0.5~ps. The positions of the surface atoms were fixed throughout the simulations. Every point of the $(a_\mathrm{fcc},E_\mathrm{ads})$ grid corresponds to a specific configuration which has been equilibrated at 290~K for 170~ns. Then 15 uncorrelated (separated by at least 10~ns) snapshots have been selected from the resulting trajectories as starting points for production runs, after having instantaneously quenched the system from 290 to 205~K. Nucleation simulations were terminated 10~ns after a significant drop of the potential energy ($> 0.53$~kcal/mol per water) was registered or if the simulation time exceeded 500~ns. In total, we report results from 6000 nucleation and 400 equilibration simulations.

The induction time $t_\mathrm{n}$ of a nucleation event has been detected by monitoring the drop in the potential energy $E_\mathrm{pot}$ of the system associated with the formation of a critical ice nucleus, as shown in Figure~\ref{FIG_2_EPOT_VS_NCLS}. We have calculated $t_\mathrm{n}$ by fitting the potential energy to:
\begin{equation} 
	E_\mathrm{pot}(t) = a + \frac{b}{1 + \exp[c(t - t_\mathrm{n})]} 
\end{equation}
where $t_\mathrm{n}, a, b$ and $c$ are fitting parameters. Due to the smoothness of the potential energy surface characterizing the mW model, crystal growth at the supercooling considered here ($\sim$70~K) is extremely fast, resulting in a very sharp potential energy drop that takes place within - at most - 1~ns for all values of $E_\mathrm{ads}$ and $a_\mathrm{fcc}$ considered. Thus, the resulting value of $t_\mathrm{n}$ does not depend on a specific functional form. We thereby estimate the error associated with the calculation of $t_\mathrm{n}$ as $\pm$ 1~ns. We also verified that no substantial discrepancy with respect to $t_\mathrm{n}$ can be observed by using other order parameters like e.g. the number $N_\mathrm{cls}$~\cite{note_orderparameter} of mW molecules in the biggest ice-like cluster, as reported in Figure~\ref{FIG_2_EPOT_VS_NCLS}.

From the $t_\mathrm{n}$ dataset, a survival probability $P_\mathrm{liq}(t)$ with respect to the metastable liquid can be built, which was then fit by a stretched/compressed exponential function:
\begin{equation}
	P_\mathrm{liq}(t) = \exp[-(J\cdot t)^\gamma]
\label{weibull}
\end{equation}
where $J$ is the nucleation rate and $\gamma$ is a parameter accounting for possible non-exponential kinetics. In fact, having quenched each starting configuration instantaneously from 290 to 205~K, we have to take into account that the relaxation of the system, when nucleation is comparably fast, could lead to a time dependent nucleation rate characterized by a non exponential behavior~\cite{sear_quantitative_2014}. Examples of $P_\mathrm{liq}(t)$ for two very different nucleation events can be found in the supporting information (SI, Figure S2).

\begin{figure}[t] 
\centerline{\includegraphics[width=7.4cm]{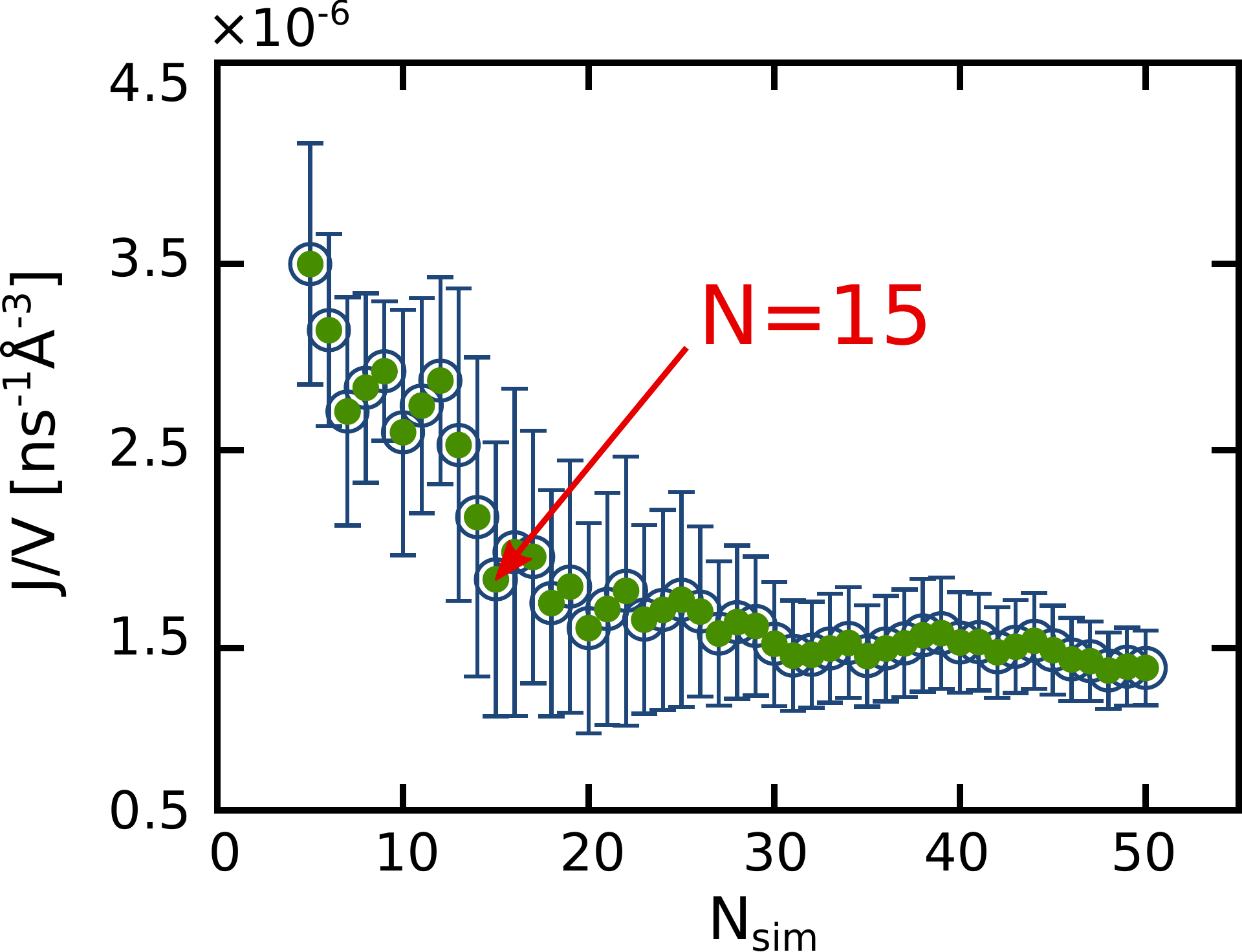}}
\caption{Illustration of the sensitivity of the nucleation rate to the number of simulations performed. Specifically the bias-corrected nucleation rate constant $J/V$ and corresponding error bars as a function of the number of simulations $N_\mathrm{sim}$, computed according to the Jackknife technique, is plotted. The data refers to the (111) surface ($E_\mathrm{ads}$=12.7~kcal/mol, $a_\mathrm{fcc}$=3.9~\r{A}).} 
\label{FIG_3_FIT} 
\end{figure}
It is difficult to quantify the error in the nucleation rates from the fitting previously described. Instead, we have employed the Jackknife resampling technique~\cite{tukey1958bias} to quantify the error associated with the finite number of simulations, and thus of induction time $t_\mathrm{n}$, that we have taken into account to compute each nucleation rate. Jackknife resampling is particularly suitable with respect to e.g. the conventional bootstrap approach when dealing with small sets of data. Results are reported in Figure~\ref{FIG_3_FIT}. The number of simulations we have chosen allows for a fairly well converged value of the nucleation rate, although an error bar accounting for about 35\% of the value has to be considered. We have chosen to estimate the error bars with respect to $J$ in the worst case scenario, namely for very mild enhancement of $J$ with respect to the homogeneous system for which very long tails in $P_\mathrm{liq}(t)$ can be observed. It must be noted that the finite size of our $t_\mathrm{n}$ dataset is the major source of error affecting the numerical accuracy of our nucleation rates. In fact, while the calculation of both $t_\mathrm{n}$ and $P_\mathrm{liq}(t)$ is basically error-free and finite size effects introduce a small systematic error, the long time tails of $P_\mathrm{liq}(t)$ can seriously suffer from a small $t_\mathrm{n}$ dataset because of the stochastic nature of the nucleation events.

\begin{figure}[t] 
\centerline{\includegraphics[width=5.8cm]{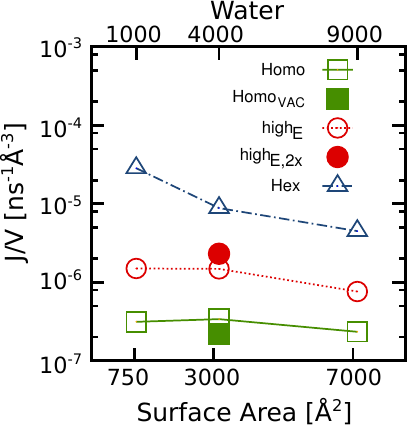}}
\caption{Nucleation rate constant $J/V$ as a function of surface area (or number of water molecules, see x-axis, top). The legend refers to bulk homogeneous nucleation (Homo), a free-standing slab (Homo$_\mathrm{VAC}$) with two vacuum interfaces, nucleation on top of the (111) surface (high$_\mathrm{E}$, $E_\mathrm{ads}$=12.7~kcal/mol, $a_\mathrm{fcc}$=3.9~\r{A}), same as high$_\mathrm{E}$ but with a water slab two times thicker (high$_\mathrm{E,2x}$), and same as high$_\mathrm{E}$ but for $E_\mathrm{ads}$=3.2 kcal/mol where we see an hexagonal overlayer (Hex).}
\label{FIG_4_FINITE_SIZE_EFFECTS}
\end{figure}

\begin{figure*}[t]
\centerline{\includegraphics[width=16cm]{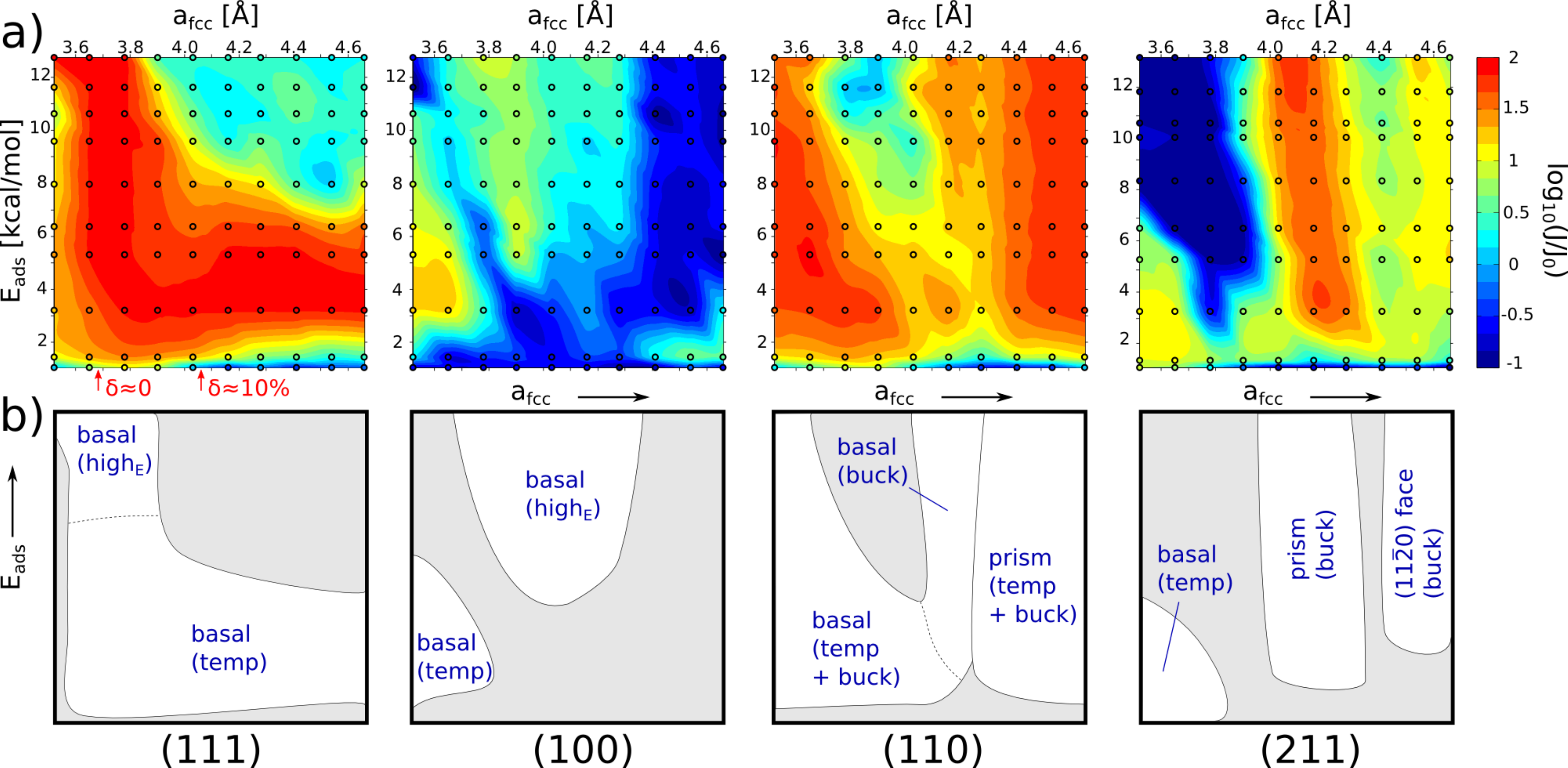}}
\caption{a) Heat maps representing the values of ice nucleation rates on top of the four different surfaces considered, plotted as a function of the adsorption energy $E_\mathrm{ads}$ and the lattice parameter $a_\mathrm{fcc}$. The lattice mismatch $\delta$ on (111) is indicated below the graph. The values of the nucleation rate $J$ are reported as $\log_{10}(J/J_0)$, where $J_0$ refers to the homogeneous nucleation rate at the same temperature. b) Sketches of the different regions (white areas) in the ($E_\mathrm{ads}$,$a_\mathrm{fcc}$) space in which we observe a significant enhancement of the nucleation rate. We label each region according to the face of $I_\mathrm{h}$ nucleating and growing on top of the surface (basal, prismatic or (11$\bar{2}$0)), together with an indication of what it is that enhances the nucleation. ``temp'', ``buck'', and ``high$_\mathrm{E}$'' refer to the in-plane template of the first overlayer, the ice-like buckling of the contact layer, and the nucleation for high adsorption energies on compact surfaces, as explained in section~\ref{sec.results.2}.}
\label{FIG_5_NUCLEATION_MAPS}
\end{figure*}

Finite size effects must be thoroughly addressed when dealing with nucleation events. At first, we have calculated the homogeneous nucleation rate $J$ as a function of volume for different models containing 1000, 4000 and 9000 mW molecules. We have considered bulk liquid models as well as free-standing water slabs, in order to take into account the influence of the vacuum-water interface that we have in our slab models. The results are summarized in Figure~\ref{FIG_4_FINITE_SIZE_EFFECTS} and led us to choose 4000 mW molecules for our heterogeneous models. Given the fact that the heterogeneous ice nucleation rates reported in this work span three orders of magnitude according to the interplay between hydrophobicity and surface morphology, we can safely state that finite size effects have little impact on our results. For instance, we have verified that doubling the area of the (111) crystalline surface (and the number of water molecules as well) only introduces a discrepancy of about a factor two in the nucleation rates (normalized by surface area) for $E_\mathrm{ads}$=3.21 or 12.76~kcal/mol ($a_\mathrm{fcc}$=3.90~\r{A}). This is somehow expected because the strong supercooling, which should guarantee a relatively small critical nucleus size. Indeed, we have obtained an estimate of the critical nucleus for a specific case ((111) surface, $E_\mathrm{ads}$=1.04~kcal/mol, $a_\mathrm{fcc}$=3.90~\r{A}) from a committor analysis~\cite{bolhuis_transition_2002} based on the number $N_\mathrm{cls}$ of mW molecules in the biggest ice-like cluster. This suggests a critical nucleus size of about only 50 mW molecules (see SI, Figure S3). This number lies consistently in the range of literature estimates, e.g. 10 molecules at 180~K~\cite{moore_is_2011} and 85~\cite{li_homogeneous_2011} to 265~\cite{reinhardt_free_2012} molecules at 220~K.

\section{Results}
\label{sec.results}
\subsection{No Simple Trend for Nucleation Rates}
\label{sec.results.1}
The nucleation rates on the four surfaces considered are shown as bi-dimensional heat maps as a function of the lattice constant and adsorption energy in Figure~\ref{FIG_5_NUCLEATION_MAPS}a. Regions in the 2D plots~\cite{note_pic} for which a strong enhancement of the nucleation rates is observed are sketched in Figure~\ref{FIG_5_NUCLEATION_MAPS}b and snapshots of representative trajectories for all the classified regions can be found in the SI (Figures S4 to S7). Before even considering any microscopic details of the water structure or nucleation processes, several general observations about the data shown in Figure~\ref{FIG_5_NUCLEATION_MAPS} can be made: 

\begin{enumerate}[leftmargin=0.3cm,itemindent=0.2cm]
\item The substrates mostly do promote nucleation compared to homogeneous nucleation. On some surfaces enhancements of up to two orders of magnitude are seen for certain values of $a_\mathrm{fcc}$ and $E_\mathrm{ads}$. The measured induction times for these events correspond to the transient time rather than the actual nucleation time, as e.g. discussed by Aga \textit{et al.}~\cite{aga_quantitative_2006} and Peng \textit{et al.}~\cite{peng_parameter-free_2010}. Therefore nucleation rates at the high end of the values reported should be seen as a lower bound rather than the actual rate.

\item Both $a_\mathrm{fcc}$ and $E_\mathrm{ads}$ do not influence nucleation on top of each surface in the same manner. Indeed, the interplay of these two parameters is different for each surface. For instance, variation in $E_\mathrm{ads}$ for the (211) surface generally has little influence on the nucleation rate. However, on the (111) surface at certain values of $a_\mathrm{fcc}$ variation in $E_\mathrm{ads}$ can have a very big impact on the nucleation rate.

\item The (111) and (110) surfaces promote ice nucleation over a much broader range than the (211) and (100) surfaces. It is worth noticing that surface symmetry alone is definitely not enough to account for such a difference. In fact, the (111) and (110) surfaces possess different symmetry (hexagonal and rectangular respectively, see Figure~\ref{FIG_1_SURFACE_LOOK}b), while the (110) and (100) surfaces, although showing completely different INA capabilities (Figure~\ref{FIG_5_NUCLEATION_MAPS}a), have quite comparable surface symmetry (rectangular and square, Figure~\ref{FIG_1_SURFACE_LOOK}b). Further evidence for the non-unique role of surface symmetry is given by the fact that simple trends are hard to find even within the very same surface. For instance, the interplay between $a_\mathrm{fcc}$ and $E_\mathrm{ads}$ in the case of the (110) surface results in two different regions where nucleation is significantly boosted (see Figure~\ref{FIG_5_NUCLEATION_MAPS}b). 

\item There is no optimal value for $E_\mathrm{ads}$. In fact it is surprising how insensitive the nucleation rate is to changes in $E_\mathrm{ads}$ for some substrates such as the (110) and (211). A notable exception is the (111) surface for $a_\mathrm{fcc} > 3.9$~\r{A}. Our results here are consistent with the recent work of Cox \textit{et al.}~\cite{cox_peeling_2015,cox_controlling_2015} where an optimal value of $E_\mathrm{ads}$ around 3 to 6~kcal/mol is found for a fcc(111) and a graphene nanoparticle. The broader range of $a_\mathrm{fcc}$ and the results for other substrates however reveal that this trend does not hold for the different morphologies.

\item A common feature on all substrates is that the nucleation rate is inhibited for the lowest value of $E_\mathrm{ads}$. For this adsorption energy the molecules basically face a hard wall which in turn could even hinder nucleation compared to the homogeneous case~\cite{reinhardt_effects_2014,haji-akbari_suppression_2014,zhang_freezing_2014,zhang_wall-induced_2014}. By analyzing the distribution of pre-critical nuclei (see SI, Figure S8) we find that these avoid the neighborhood of the surface for mentioned $E_\mathrm{ads}$ range. The effect therefore can be roughly rationalized in a smaller volume available for the nuclei to appear. This volume can be estimated by the area affected by significant density perturbations due to the presence of the surface (see SI, Figure S9). This kind of inhibition is unlikely to be visible in simulations or experiments where the ratio of water volume to contact area is much higher than in our case.

\item The lattice mismatch $\delta$ cannot be regarded as a requirement for an INA. This issue specifically concerns the (111) substrate, because of its compatibility with the basal face of hexagonal ice $I_\mathrm{h}$. We calculated $\delta$ according to equation \ref{eqn.mismatch}, the lattice constant of ice $a_\mathrm{i}\approx 4.51$~\r{A}~\cite{pruppacher1997microphysics} and $a_\mathrm{s} = \sqrt{3/2}\cdot a_\mathrm{fcc}$. Therefore a value of $a_\mathrm{fcc} = 3.68$~\r{A} corresponds to a zero mismatch ($\delta = 0$) which is indicated in Figure~\ref{FIG_5_NUCLEATION_MAPS}a. If $\delta \approx 0$ is the main requirement for enhanced ice nucleation, we would expect a distinct peak around the corresponding value of $a_\mathrm{fcc}$. The results for nucleation rates however clearly show that this is not the case. Certainly, a small value of $\delta$ does promote nucleation for a wider range of $E_\mathrm{ads}$, but for adsorption energies between 2 and 6~kcal/mol enhanced nucleation is observed for mismatches even beyond +10~\%. We note that for $\delta < 0$ the drop of nucleation rates seems to start sooner than for $\delta > 0$, although the corresponding lattice constants lie somewhat outside of our considered range. This is consistent with all atom simulations which show that a mismatch $\delta$ slightly larger than $0$ is more favorable~\cite{cox_non-hexagonal_2012}. Furthermore, Mithen and Sear~\cite{mithen_computer_2014} computed nucleation rates of a Lennard-Jones liquid in contact with a substrate and found the maximum close to, but larger than $\delta = 0$. Overall, our results suggest that a small lattice mismatch is helpful to nucleation, but cannot be regarded as the most important requirement for an INA. For the other surfaces, the definition of disregistry $\delta$ is not as straightforward, because the substrates do not provide a clear template. In fact a strict definition of what can be regarded as similar or not similar is not part of the lattice mismatch theory. We have therefore restricted our discussion of the lattice mismatch to the (111) substrate.

\end{enumerate}

\subsection{Microscopic Factors for Nucleation}
\label{sec.results.2}
It is somehow unexpected that a simplistic model like the one used here can foster such diverse behavior. However, when we examine the water structures and nucleation processes in detail, general trends do emerge. We now discuss the key features important to nucleation.
\subsubsection{In-Plane Template of the First Overlayer}
\begin{figure}[t]
\centerline{\includegraphics[width=8cm]{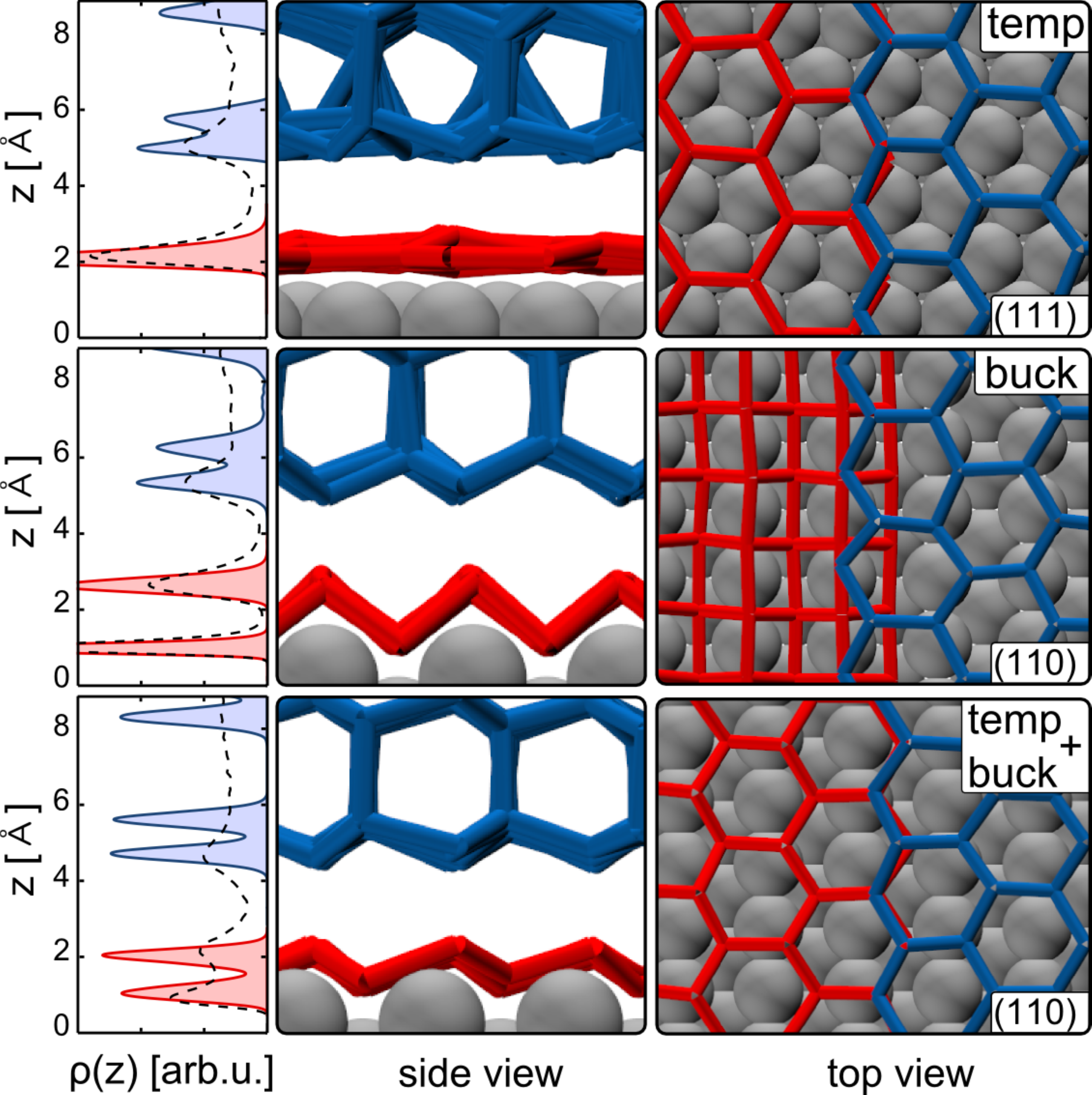}}
\caption{Analysis of certain factors important to nucleation. Each row represents data obtained from a representative trajectory for events classified as ``temp'', ``temp'' and both combined (``temp + buck'') mechanisms (see section~\ref{sec.results.2}). The first column depicts the density of water molecules above the surface after freezing (filled curves) and during equilibration before freezing (dashed black line). The second column shows side views and the third column snapshots viewed from above. In all cases the contact layer is colored red while higher layers are colored blue. For ease of visualization in the top view only part of the second layer is shown.}
\label{FIG_6_DENSITIES_AND_LAYERING}
\end{figure}

The in-plane structure of the first water overlayer plays an important role in nucleation, because it can act as a template to higher layers. This is particularly evident on the (111) surface, which possesses an hexagonal symmetry compatible with the in-plane symmetry of the basal face of ice (honeycomb). Where nucleation is significantly enhanced, we find that an hexagonal overlayer (HOL) of water molecules forms on top of the surface (Figure~\ref{FIG_6_DENSITIES_AND_LAYERING}, ``temp''), rapidly inducing nucleation. The promotion-effect persists even when a significant number of defects, such as 4-, 5- or 7-membered rings appear within the HOL, as well as in the case of larger lattice mismatches $\delta > 0$ where the HOL is severely stretched. This indicates that the template does not have to be perfect to promote nucleation. The HOL rules the majority of nucleation processes on top of the (111) surface, where only the basal face has been observed to nucleate and grow (see Figure~\ref{FIG_5_NUCLEATION_MAPS}b). However, contrary to the idealized bilayer structure of the basal face the overlayers observed here are mostly flat. Reduced buckling in the contact layer has been suggested in a number of studies on metals~\cite{ogasawara_structure_2002,michaelides_insight_2004}. The flat hexagonal structures identified here which precede nucleation indicate that a good template needs: (i) the right symmetry and (ii) the right intermolecular distances in the plane, but not necessarily the correct water molecule heights.

We have labeled nucleation events induced by this contact layer as ``temp'' rather than ``hex'' to stress that it is not exclusively the bi-layer template of the basal face, typically associated with the term HOL, but rather any possible overlayer compatible with a face of ice. An example of a different overlayer is found on the (110) surface (see SI, Figure S7) compatible with the prismatic face of ice. 

\subsubsection{Buckling of the First Overlayer}
Our results concerning the (110) and (211) surfaces suggest that different heights of atoms in the contact layer, termed buckling, is an important factor to enhanced nucleation. The difference between a flat and a buckled overlayer can be seen in the water density and the side views of selected trajectories, depicted in Figure~\ref{FIG_6_DENSITIES_AND_LAYERING}. The density for an event characterized by the ``temp'' mechanism has only a single spike representing the flat hexagonal contact layer. In contrast on the (110) substrate at large lattice constants (Figure~\ref{FIG_6_DENSITIES_AND_LAYERING}, ``buck''), the first water overlayer is not ice-like but exhibits a pronounced buckling of the contact layer. The fact that we find this combination of a symmetrically unfavorable (and therefore non-templating) but buckled contact layer for many of the enhanced nucleation trajectories leads us to conclude that the buckling in this case is the microscopic cause for the nucleation enhancement (labeled as ``buck''). 

\begin{figure*}[t]
\centerline{\includegraphics[width=15.4cm]{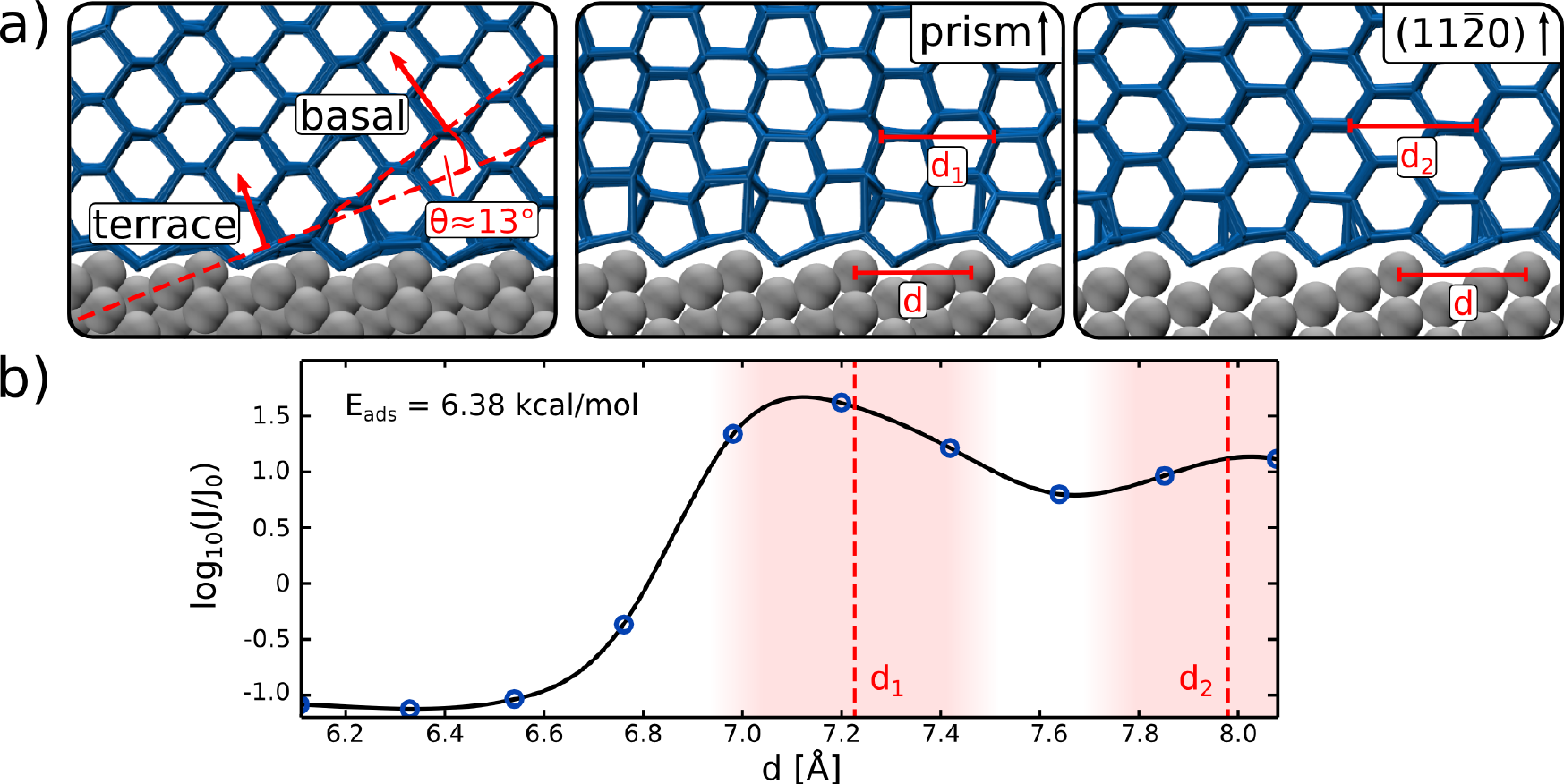}}
\caption{(a) Representative snapshots of the three different faces (basal, prismatic and (11$\bar{2}$0) face) of hexagonal ice growing on top of the (211) surface (side view). Surface atoms are depicted as balls (grey), while the bonding network of water molecules is represented by sticks (blue). The $\theta$ angle in the top left panel illustrates that the basal face and the normal of the (111) terrace deviate.
b) Nucleation rates (circles) and spline interpolation (line) on the (211) surface as a function of the step distance $d$. The red lines indicate the measured characteristic distances $d_1$ and $d_2$ as well as their standard deviation (red shaded area). The meaning of $d$, $d_1$ and $d_2$ is illustrated in the top panels.}
\label{FIG_7_DIFFERENT_FACES}
\end{figure*}

As with the (110), the (211) geometry also leads to a significant enhancement of nucleation rates in specific regions. In addition, and quite surprisingly on this surface, nucleation and growth of three different faces of ice are observed. The three regimes roughly correspond to different values of $a_\mathrm{fcc}$ (Figure~\ref{FIG_7_DIFFERENT_FACES}a). The (211) substrate has a rectangular in-plane symmetry, but it features (111) micro-facets (see Figure~\ref{FIG_1_SURFACE_LOOK}a). For small values of $a_\mathrm{fcc}$ (Figure~\ref{FIG_7_DIFFERENT_FACES}a), the spacing between the steps allows for rows of hexagons to form on top of these terraces. This template has a symmetry consistent with the basal face of $I_\mathrm{h}$ which in fact exclusively nucleates in this first regime. As an aside we note that the growth direction of the basal face is not exactly parallel to the surface normal of the (111) terraces, leading to the small angle mismatch shown in Figure~\ref{FIG_7_DIFFERENT_FACES}a. As we move on to larger lattice constants, the spacing between the steps becomes too large to accommodate an hexagonal overlayer. Rather a rectangular overlayer appears on top of the surface, wiping out the templating effect of the hexagons. These overlayers are buckled in a manner that follows the corrugation of the surface. This results in the nucleation and growth of the prismatic and ($11\bar{2}0$) faces for $a_\mathrm{fcc}$ values of $4.16$~\r{A} and $4.66$~\r{A} respectively (see Figure~\ref{FIG_7_DIFFERENT_FACES}a). The contact layers, despite being significantly buckled, do not show a favorable in-plane template (pictures of the overlayers can be found in the SI, Figure S5). 

In the specific case of the (211) substrate $a_\mathrm{fcc}$ is a much more sensitive parameter for the nucleation rate than $E_\mathrm{ads}$ which only leads to changes for vanishing interaction. This suggests that nucleation enhancement by the buckling of the overlayer is a more geometrical phenomenon. Indeed, we find that the buckling of the contact layer in these cases coincides with a characteristic periodic length of one of the ice faces. To support this interpretation, in Figure~\ref{FIG_7_DIFFERENT_FACES}b the nucleation rates for $E_\mathrm{ads}=6.38$~kcal/mol are displayed as a function of the step distance $d$. The characteristic lengths for prism and ($11\bar{2}0$) face ($d_1$ and $d_2$ respectively) were obtained from measuring and averaging the corresponding distances in representative trajectories where we see freezing of that particular ice face. The values correspond well with the step periodicity $d$ at which nucleation is enhanced the most through formation of the respective face. A similar conclusion was drawn by Zhang \textit{et al.}~\cite{zhang_impact_2014} for trenches promoting nucleation the most when they resemble a characteristic spacing. These effects seem to fade when the roughness is on a larger than atomic scale~\cite{campbell_is_2015} or if the surface is amorphous~\cite{lupi_heterogeneous_2014}. 

The results shown in Figure~\ref{FIG_7_DIFFERENT_FACES}b are reminiscent of the predictions of Turnbull and Vonnegut~\cite{turnbull_nucleation_1952} regarding a small lattice mismatch. Indeed, if one neglects the fact that the atomic arrangements of the substrate and respective ice face at the interface are dissimilar, the buckling can be interpreted as a lattice mismatch. However, this concept is unlikely to be helpful in general as it does not clearly distinguish the two ingredients that form the buckling: (i) the different heights of atoms that are adsorbed onto the surface and; (ii) the periodicity that describes the variation of atomic heights. Contrary to the lattice mismatch, a compatible in-plane template is not required for the buckling.

We also observed nucleation events in which the overlayer possessed both atomic scale buckling and favorable in-plane template. In two specific $E_\mathrm{ads}$ and $a_\mathrm{fcc}$ intervals a buckled first overlayer displaying an in-plane template consistent with the basal or prismatic face respectively (labeled as ``temp + buck'') forms on the (110) surface. The third row in Figure~\ref{FIG_6_DENSITIES_AND_LAYERING} depicts the density and water overlayers in the case of basal face growth. Here, the structured water during equilibration exhibits an appearance that is already close to the double-peak of frozen (basal face) ice. The resulting overlayer consists of hexagonal arrangements, comparable to the basal face of ice - not only as an in-plane template, but also in the buckling. The importance of ice-like structuring along the z direction has been observed and discussed in the case of AgI~\cite{zielke_molecular_2014}. Notably an HOL is not enough in this case, suggesting that ice-like buckling could be more effective than in-plane templating.

On the other hand, when neither the in-plane template or the favorable buckling are present, no sizable enhancement of the nucleation rate has been observed. This is what happens for the majority of the ($E_\mathrm{ads}$,$a_\mathrm{fcc}$) points on the (100) surface (see Figure~\ref{FIG_5_NUCLEATION_MAPS}), which has a square symmetry and being perfectly flat does not cause the contact layer to buckle.

\subsubsection{High adsorption-energy nucleation on compact surfaces}
We have also observed the promotion of nucleation in two regions where neither the ice-like in-plane template or buckling of the contact layer was present. The two regions can be found for the (111) and (100) surfaces (see Figure~\ref{FIG_5_NUCLEATION_MAPS}) and have been labeled high$_\mathrm{E}$ to emphasize that they occur only for the higher adsorption energies. It is also apparent that we find this kind of enhancement on the two compact surfaces rather than the more open ones, which suggests that it is the combination of strong interaction and surface denseness that facilitates the nucleation. The overlayers in these cases were very dense (a disordered overlayer for (111) and perfect squares for (100), see the SI Figures S4 and S6). It is clear that these structures should be anything but advantageous for nucleation. The analysis of the distribution of pre-critical nuclei for a representative point (see SI, Figure S8) reveals that nucleation happened on top of the first 2$\sim$3 water layers. It is therefore clearly a heterogeneous event which the increased rates already suggested. While the actual reason for this kind of nucleation enhancement is not immediately obvious and potentially interesting, it must be noted that values of $E_\mathrm{ads}$ in the upper third of the considered range are abstract, as water will probably dissociate on top of the surface rather than being adsorbed. Thus, we have not made further investigations concerning this specific enhancement, however we suggest two possible effects that could be the driving force behind it. First, a layering mechanism similar to the one discussed by Cox \textit{et al.}~\cite{cox_peeling_2015} could influence higher layers when the coverage of the underlying layers is saturated. This is also supported by the values of layering we have calculated, as discussed later (\ref{sec.results.3}). A second reason for the facilitation could also be dynamical effects, which have been shown to significantly influence molecules and atoms near the interface~\cite{haji-akbari_effect_2014}. The strong adsorption causes the first 1$\sim$2 layers to be nearly immobile, effectively extending the surface height and possibly shifting the dynamical effects to layers above $\sim$10~\r{A}. Lastly, we note that the effects of high$_\mathrm{E}$ nucleation could be shifted towards more realistic interactions for all atom-models of water, since in our tests with the TIP4P/2005 model we observed a slightly more pronounced structuring and layering (see SI, Figure S10).    
\begin{figure*}[t]
\centerline{\includegraphics[width=15.9cm]{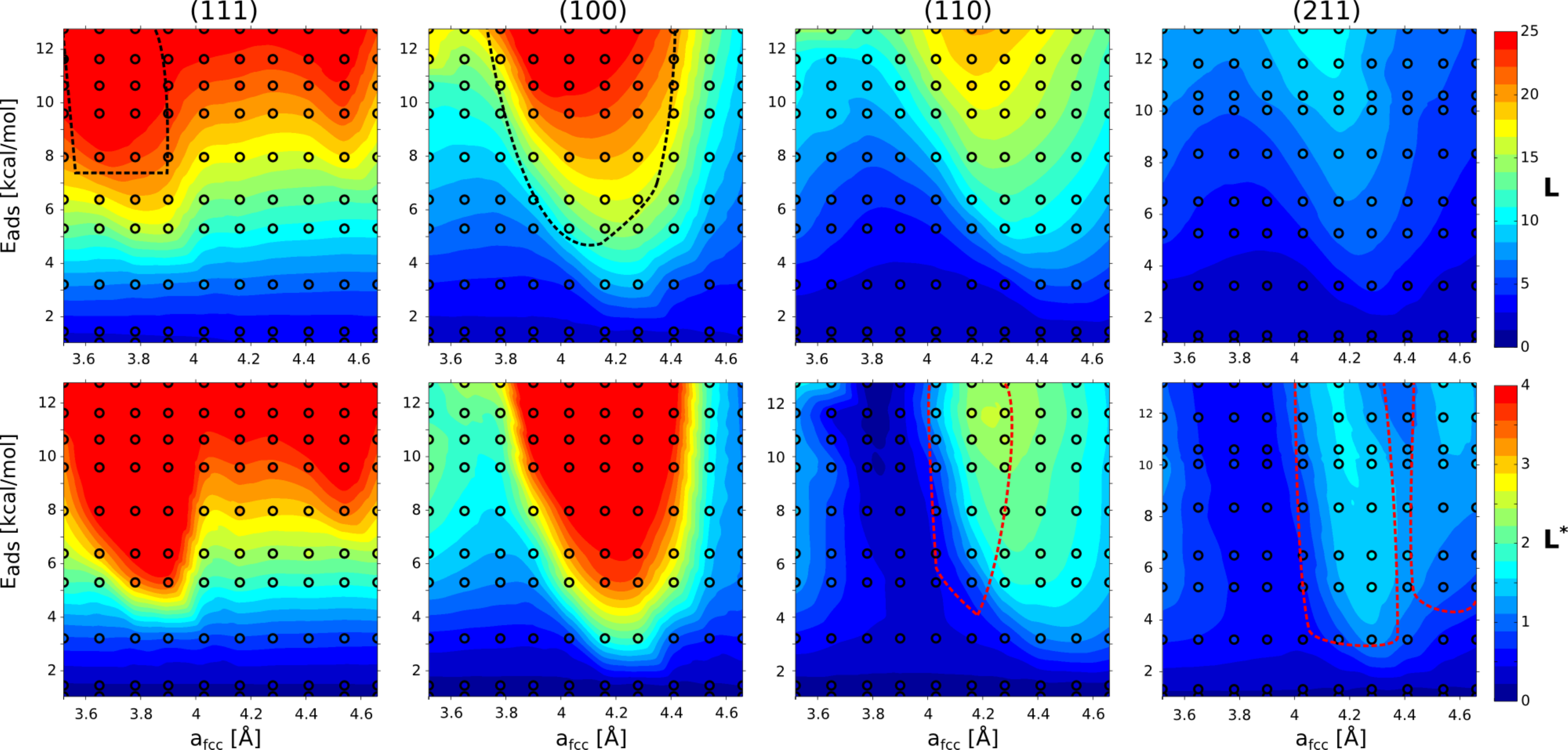}} 
\caption{Heat maps representing the total layering $L$ (top) and the layering excluding the contact layer $L^*$ (bottom) calculated from the equilibration runs. The dashed lines indicate the regions where nucleation was enhanced through high$_\mathrm{E}$ nucleation (black) and buckling (red). Note that the color range for $L$ and $L^*$ is different.} 
\label{FIG_8_LAYERING} 
\end{figure*}

\subsection{Further Insight and Future Perspective}
\label{sec.results.3}
Having examined heterogeneous ice nucleation on the four crystalline substrates and identified some of the key factors responsible for the enhanced nucleation observed, we now discuss a number of open issues and ways this work could be taken forward in the future. 

\subsubsection{Layering}
Lupi \textit{et al.}~\cite{lupi_does_2014,lupi_heterogeneous_2014} found that the layering $L$ of water at graphitic interfaces correlates with their freezing efficiency. For high adsorption energies Cox \textit{et al.}~\cite{cox_peeling_2015} also found a correlation between the nucleation and the layering, but only if the layering associated with the contact layer was excluded ($L^*$). These two forms of layering are defined as follows:
\begin{equation}
	L = \int_{0}^{z_\mathrm{bulk}} \left|\frac{\rho(z)}{\rho_0}-1\right|^2 \mathrm{d}z
\end{equation}
\begin{equation}
	L^* = \int_{z_0}^{z_\mathrm{bulk}} \left|\frac{\rho(z)}{\rho_0}-1\right|^2 \mathrm{d}z
\end{equation}
where $\rho_0$ is the density of bulk water, $z_\mathrm{bulk}$ is the height above the surface at which the water density $\rho$ converges to the bulk value and $z_0$ is a height so that the layering contributions of the contact layer are excluded. In all cases $z_\mathrm{bulk}=18$~\r{A} was used. The results are shown in Figure~\ref{FIG_8_LAYERING}.

Generally, a comparison between the layering plots and the nucleation rates in Figure~\ref{FIG_5_NUCLEATION_MAPS} shows that $L$ and $L^*$ do not correlate very well with the nucleation ability of the surface. We find that both $L$ and $L^*$ monotonically increase with $E_\mathrm{ads}$. However, there seems to be a non-trivial dependency on $a_\mathrm{fcc}$ where for 3 of the 4 surfaces the layering first increases towards medium values of the lattice constant and then decreases again. This trend reflects a change of the adsorption structure of the water molecules, which was also partially the cause of the different mechanisms influencing nucleation rates. However, the trends of the nucleation maps are not reproduced. For instance, no region that has been classified as ``temp'' can be distinguished from its surrounding in the layering plots. If one assumes an optimal value or threshold for $L$ or $L^*$, the corresponding iso-surfaces in the plot would not reproduce any plot of the nucleation rates. While this conclusion has already started to emerge from the work of Cox \textit{et al.}~\cite{cox_peeling_2015} we can now base the argumentation on a much larger parameter space and additional surfaces.

Interestingly, we find some exceptions. Specifically the areas classified as high$_\mathrm{E}$ nucleation seem to be similar to areas of strong layering. This could also explain why we do not see this mechanism on the (110) and (211) surfaces, because the layering is still too weak even for the highest adsorption energies. Also the $L^*$ plot for the (211) substrate seems to indicate the regions that have been classified as ``buck''. However no quantitative agreement is found, as for instance $L^* \approx 2$ on the (110) surface shapes a region where nucleation rates differ by an order of magnitude.

Overall, we find that the layering does not generally correlate with the nucleation ability which is likely due to the fact that this quantity averages over lateral structure effects such as the in-plane symmetry and template. The exceptions are such cases where the potential energy surface is smooth, as for instance high$_\mathrm{E}$ events where nucleation happens further away from the surface or nucleation on graphene-like surfaces~\cite{lupi_does_2014,cox_peeling_2015}. 

\subsubsection{Notes on the water model}
The fast dynamics associated with the coarse grained mW model has made the current systematic study possible. However it is important to consider, at the very least, how the absence of explicit hydrogens affects the results of this study. To this end we have performed test calculations with the all-atom TIP4P/2005 model~\cite{abascal_general_2005} which provides a reasonable description of water~\cite{vega_simulating_2011}.

Firstly, we compared the water densities for one ($a_\mathrm{fcc}$,$E_\mathrm{ads}$) point on each surface (see SI, Figure S10). The densities obtained are very similar for both water models and we conclude that the structuring they show is nearly equivalent.

Secondly, a set of nucleation simulations with the all-atom model was carried out on the (111) surface. Here, the almost instantaneous formation of an hexagonal overlayer was the driving force behind the nucleation enhancement for mW. With TIP4P/2005 we did not observe the complete formation of such an overlayer within $100$~ns. However, an analysis of the hexagonal cluster distribution (see SI, Figure S11) shows that the largest patches of hexagons can be found for $E_\mathrm{ads}\approx 3.2$~kcal/mol. This is precisely the value for which we observe the fastest formation in the case of mW as well~\cite{note_heatofvap}. This trend confirms that while - especially heterogeneous - nucleation processes modeled by mW water are for certain nonphysically fast, they can still capture part of the underlying physics.   
 
\subsubsection{Higher Temperatures}
It is interesting to understand how the trends observed in this study could depend on temperature, especially because our simulations were performed in the deeply supercooled regime. It is currently beyond reach to carry out such an extensive set of simulations at a significantly higher temperature with the brute-force approach. However, to estimate the effect of the strong supercooling we performed further calculations at 210 and 215~K for 3 adsorption energies on the (110) surface (the results can be found in the SI, Figure S12). We find that the nucleation rates in regions where no specific mechanism has been attributed heavily decline, but otherwise no significant changes can be observed. That includes the trends of the nucleation rate as well as the adsorption structures, which are the basis for the mechanisms we propose. This indicates that our conclusions are also valid for higher temperatures.

\subsubsection{Future Perspective and Experimental Verification}
Before concluding we discuss some aspects that should be addressed in future studies as well as making some suggestions about how the insight presented here could be tested experimentally. 

A first step will be to use all-atom models of water~\cite{vega_simulating_2011,sanz_homogeneous_2013} such as the TIP4P/2005 discussed above or its cousin TIP4P/Ice~\cite{abascal_potential_2005} specifically designed for the study of ice. Recently all-atom simulations of homogeneous ice nucleation have been performed~\cite{haji-akbari_direct_2015} with the help of the forward-flux sampling technique~\cite{allen_simulating_2006,allen_forward_2009}. The latter seems like a promising approach for nucleation simulations~\cite{li_homogeneous_2011,li_ice_2013,jiang_efficient_2013,bi_probing_2014,haji-akbari_suppression_2014,cabriolu_ice_2015,haji-akbari_direct_2015} although there are of course many other free energy and enhanced sampling techniques~\cite{torrie_nonphysical_1977,mezei_adaptive_1987,laio_escaping_2002,bolhuis_transition_2002,laio_metadynamics:_2008,kastner_umbrella_2011,zheng_rapid_2013,dittmar_driving_2014,tiwary_time-independent_2015} that could be used. Improvement in the water-surface interaction potential is of equal importance if one wishes to investigate heterogeneous ice nucleation. For instance, an extension of the present study to realistic clean metal surfaces needs to account for the orientational dependence of the water molecules on the surface and polarization effects. Fitting water-surface interaction potentials to density functional theory or higher-level electronic structure theories is one way to take such effects into account and work in our group in this direction is ongoing~\cite{ma_adsorption_2011,carrasco_role_2013,al-hamdani_water_2014,al-hamdani_communication:_2015}. Furthermore, it has been shown that dissociation of water molecules occurs at reactive metal surfaces so that the overlayers can be comprised of water-hydroxyl mixtures~\cite{michaelides_catalytic_2001,feibelman_partial_2002,forster_2011,carrasco_molecular_2012}. Taking this issue into account will require a suitable and accurate dissociable model of water. Lastly, it will be important that nucleation studies approach experiments more closely. Especially the supercooling in computational studies is a major concern since it is too strong to directly allow for comparison with e.g. atmospheric or laboratory measurements. 

Our results could be most directly probed by measurements that can reliably characterize surface structures with molecular level of resolution. This would require ultra-high vacuum prepared levels of cleanliness. A most promising candidate for an experimental study would be gold surfaces because of their resistance to oxidation and golds fcc crystal structure. With $a_\mathrm{fcc} \approx 4.08$~\r{A}~\cite{wolfram} and $E_\mathrm{ads} \approx 3$-$7$~kcal/mol~\cite{carrasco_role_2013} our simulations indicate that the nucleation rates on the (111) and (100) gold substrates should differ by $2$-$3$ orders of magnitude. This has been estimated from the data in Figure~\ref{FIG_5_NUCLEATION_MAPS} in the region of the Au lattice constant and $E_\mathrm{ads}$~\cite{note_largerDifference}. This would also allow the control of nucleation on gold (and other) nano-particles that expose different facets. By adding molecules that are inactive for ice nucleation but selectively bind to the promoting facets of the particle the nucleation rate could be controlled. Indications of freezing in a well defined surface-science-style study could be obtained with e.g. ambient pressure x-ray photoelectron spectroscopy~\cite{bluhm_premelting_2002,ketteler_nature_2007} or surface x-ray diffraction~\cite{gustafson_high-energy_2014}. Another class of interesting materials are halogenated graphene~\cite{di_stability_2011,georgakilas_functionalization_2012} and graphane~\cite{elias_control_2009}. The functionalization of graphene with different atoms such as H, F, Cl, Br or I should alter the underlying geometry of the 2D material only slightly~\cite{medeiros_dft_2010,gao_density_2011}, but the water-surface interaction could greatly vary~\cite{garcia_group_2011}. This could be exploited to verify our predictions for different interaction strengths by examining ice nucleation on these compounds. This idea can even be extended to other quasi-2D honeycomb materials such as silicene~\cite{vogt_silicene:_2012}, germanene~\cite{davila_germanene:_2014} and stanene~\cite{zhu_epitaxial_2015} that have different lattice constants~\cite{garcia_group_2011,wei_many-body_2013} if grown on appropriate supports and if they remain stable in an aqueous environment. In such a manner the interplay between morphology and hydrophobicity could be examined experimentally, possibly yielding a similar nucleation map to Figure~\ref{FIG_5_NUCLEATION_MAPS}a. Moreover, self assembled monolayers~\cite{seeley_two-dimensional_2001,arnold_preparation_2002,ochshorn_towards_2006} provide the possibility to create specific morphologies. For instance different headgroups for aliphatic chains can alter the hydrophobicity of the resulting surface, while functional groups in the chain can change the spacing between them. Additionally, different chain lengths could be used to design a buckled surface. In combination with non hydrogen-bonding headgroups this could enable the design of interfaces useful for testing the buckling mechanism. Finally, we note that the exciting capabilities of femtosecond x-ray scattering~\cite{sellberg_ultrafast_2014,laksmono_anomalous_2015} techniques that have recently been used to explore homogeneous ice formation in water droplets could possibly be extended to heterogeneous systems.



\section{Conclusions}
\label{sec.conclusions}
In summary, we have examined the interplay between surface morphology and hydrophobicity on the ability of a generic crystalline surface to promote ice nucleation. We have calculated the nucleation rates of a coarse grained model of water on top of four different crystalline surfaces of an ideal fcc crystal by means of brute-force molecular dynamics simulations, sweeping a comprehensive range of adsorption energies and lattice parameters. 

Strikingly different nucleation scenarios have emerged on the various crystalline surfaces considered. Even for a specific surface the balance between lattice constant and hydrophobicity fosters non trivial trends. Most surprisingly the nucleation and growth of up to three different faces of hexagonal ice on top of the same surface could be induced by altering the lattice parameter alone. 

We have demonstrated that on the (111) surface a small lattice mismatch  with respect to ice is certainly not a requirement for promoting ice nucleation. This implies that in the search for understanding of the nucleation performance of known materials or the design of new ones one should not exclusively focus on the lattice mismatch issue. Nonetheless, our results show that it is important which surface is present, as nucleation rates can vary from inhibition to promotion for different faces of the same material. This means that experiments have to carefully characterize the atomic structure of INAs, because the sheer morphological difference in samples could account for varying nucleation rates. Additionally, this provides exciting possibilities to change the ice nucleation behavior of materials through e.g. growth-habit control~\cite{Hansen_atom_2002} to strengthen the inhibition effect or to turn nucleating nano particles into inhibitors and \textit{vice versa}. 

In most cases nucleation is promoted in a wide range of $E_\mathrm{ads}$ without changing the molecular mechanism. Therefore, optimal interaction strengths are rare exceptions and only found for some specific $a_\mathrm{fcc}$ ranges.  

Finally, we have pinpointed three different scenarios that facilitate the nucleation process. 
\begin{enumerate}
	\item The ability of the surface to create a first water overlayer that provides an in-plane template consistent with one of the faces of ice. Such an overlayer is typically, but not exclusively found on top of a surface that already displays a compatible symmetry.
    \item The ability of the surface to structure the first two water overlayers in such a way that they resemble either the density profile perpendicular to the surface or a characteristic buckling distance in the surface plane of one of the faces of ice. This typically requires a certain roughness at the atomic scale. 
    \item Even in the case of a first overlayer lacking both an in-plane template and structuring, nucleation can be promoted within the higher water layers. This kind of enhancement requires a compact surface with high adsorption energy. 
\end{enumerate}

Whether or not one of these scenarios could take place on top of a given surface, depends in a non trivial manner on both the morphology and hydrophobicity of the surface. Such a large body of findings will hopefully encourage and guide future work addressing heterogeneous ice nucleation on top of realistic surfaces, in the hope of furthering our understanding of what makes a material a good ice nucleating agent. 

\section*{Acknowledgement}
This work was supported by the European Research Council under the European Union's Seventh Framework Programme (FP/2007-2013) / ERC Grant Agreement number 616121 (HeteroIce project). A.M. is also supported by the  Royal Society through a Royal Society Wolfson Research Merit Award. Dr. A. Zen and P. Pedevilla are thanked for insightful discussions. We thank Dr. Ben Slater for commenting on the manuscript. We are grateful for computational resources provided by the London Centre for Nanotechnology and the Materials Chemistry Consortium through the EPSRC grant number EP/L000202. 

\input{paper.bbl}

\clearpage
\onecolumngrid 
\setcounter{section}{0}
\renewcommand{\thesection}{S\arabic{section}}%
\setcounter{table}{0}
\renewcommand{\thetable}{S\arabic{table}}%
\setcounter{figure}{0}
\renewcommand{\thefigure}{S\arabic{figure}}%
\section*{Supporting Information}
\raggedbottom


\subsection{Distribution of the $\mathrm{\bar{q}_3(i)}$ Order Parameter}
\begin{figure}[h]
\centerline{\includegraphics[width=11cm]{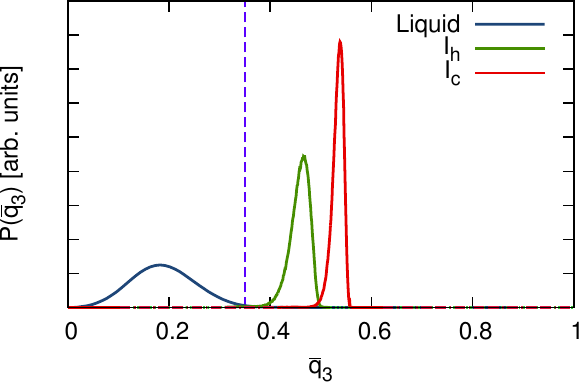}} 
\caption{Distribution for the $\bar{q}_3(i)$ parameter for liquid water, hexagonal ice $I_\mathrm{h}$ and cubic ice $I_\mathrm{c}$. The data was obtained by a short time simulation of the liquid and pristine crystals respectively (205~K, 4096 molecules, NPT, 10 ns). The dashed line indicates the threshold above which particles have been considered as solid.}
\label{FIG_SI_Q3} 
\end{figure}

Nucleation was monitored by following the change in the potential energy. As a separate check, for selected trajectories we also monitored $N_\mathrm{cls}$, the number of molecules in the largest solid-like cluster. The state of molecules was characterized by a modified version of the local $\bar{q}_3(i)$ parameter~\cite{lechner_accurate_2008}. Figure~\ref{FIG_SI_Q3} shows the distribution of $\bar{q}_3(i)$ for different phases. We applied a cutoff of $3.2$~\r{A} for both the $\bar{q}_3(i)$ neighbor-list and the cluster algorithm.

\subsection{Compressed Exponential Fit}
\begin{figure}[h]
\centerline{\includegraphics[width=12cm]{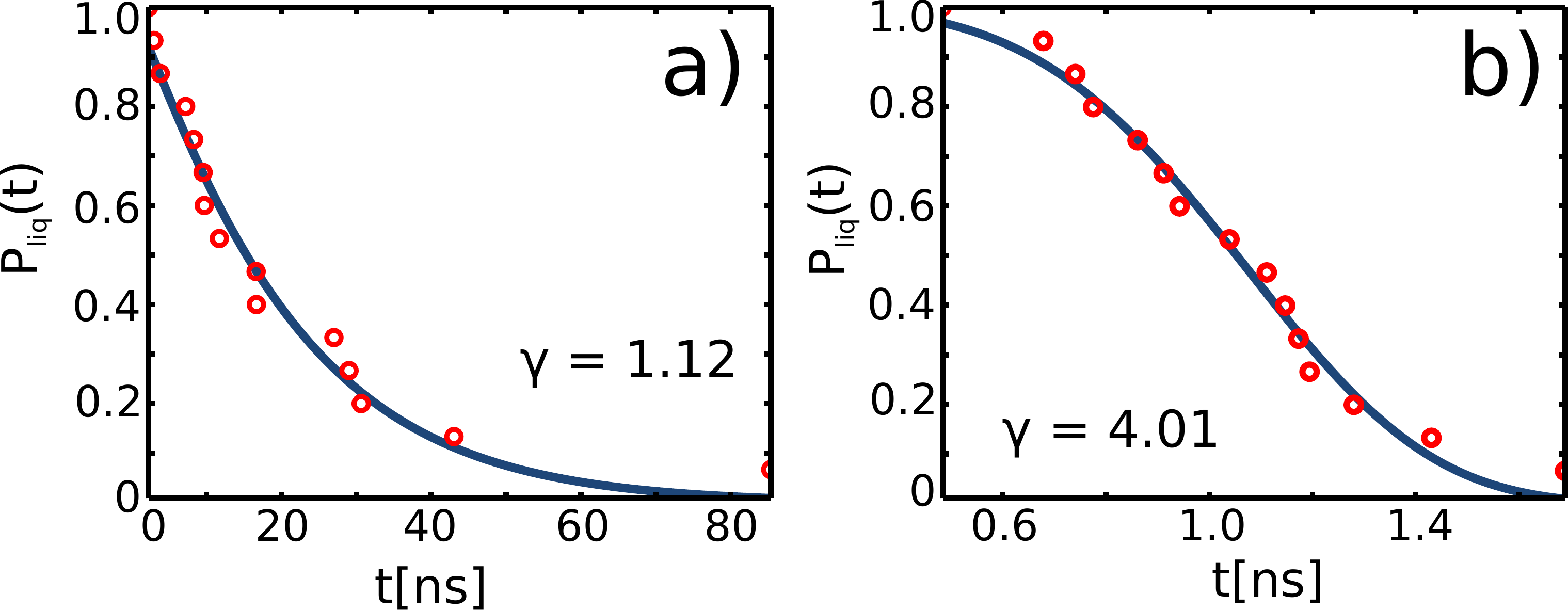}}
\caption{Compressed exponential fitting results for two dissimilar nucleation events. $P_\mathrm{liq}(t)$ (red circles) and fit after equation 4 (blue lines) for the (110) surface and $a_\mathrm{fcc}=3.9$~\r{A}. a) $E_\mathrm{ads}=11.63$~kcal/mol and b) $E_\mathrm{ads}=5.3$~kcal/mol.} 
\label{FIG_SI_STRETCHED_EXPONENTIAL} 
\end{figure}
The simulation protocol involves an instantaneous quench from the equilibration temperature to the one at which we study nucleation. Because the system has to relax into quasi-equilibrium first, the nucleation rate increases with time, resulting in a deviation from perfect exponential characteristics. The effect of this non-exponential behavior can be appreciated in Figure~\ref{FIG_SI_STRETCHED_EXPONENTIAL}, where we show the $t_\mathrm{n}$ datasets and the resulting $P_\mathrm{liq}(t)$ for two dissimilar nucleation scenarios observed on the (110) surface as a function of the strength of the water-surface interaction. In the case of a) the nucleation typically proceeds on a timescale ranging from 1 to 100 ns, resulting in well behaved exponential decay ($\gamma\sim$1 in equation 4 for the survival probability). On the other hand, the fitting of the data shown in Figure~\ref{FIG_SI_STRETCHED_EXPONENTIAL}b gave $\gamma \gg 1$, which in turn implies a nucleation rate that increases with time, as the timescale for $t_\mathrm{n}$ (0.1-1~ns) is indeed comparable with the relaxation time of the system. This occurrence takes place mainly for those ($E_\mathrm{ads}$,$a_\mathrm{fcc}$) values for which we observe the basically instantaneous (10-1000~ps) formation of almost perfect ice-like overlayers on top of the surface.

\subsection{Critical Nucleus Size on the (111) surface}
\begin{figure}[h]
\centerline{\includegraphics[width=9cm]{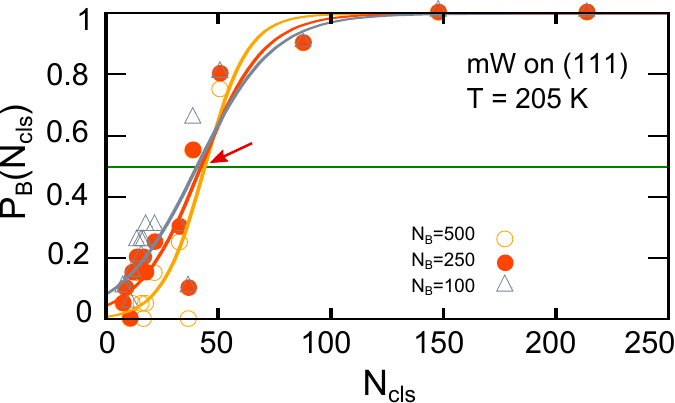}} 
\caption{Committor probability ($P_{B}$) with respect to the number $N_\mathrm{cls}$ of mW molecules in the biggest ice-like cluster for the (111) surface ($E_\mathrm{ads}=1.04$~kcal/mol, $a_\mathrm{fcc}=3.90$~\r{A}). Three different thresholds $N_\mathrm{B}$ for the order parameter have been considered and reported. The analysis has been obtained by shooting 30 statistically independent MD runs (2 ns long) from 40 different starting configurations taken along a nucleation trajectory. The arrow marks the critical nucleus size $\approx 50$.}
\label{FIG_SI_NUCLEUS} 
\end{figure}

To obtain an estimate of the critical nucleus size we performed a committor analysis~\cite{bolhuis_transition_2002} on the (111) surface. The results are depicted in Figure~\ref{FIG_SI_NUCLEUS} and suggest a critical nucleus size of circa 50 mW molecules, which is much smaller than our system size (4000 mW).

\subsection{Snapshots of Classified Regions}
\begin{figure}[H]
\centerline{\includegraphics[width=10cm]{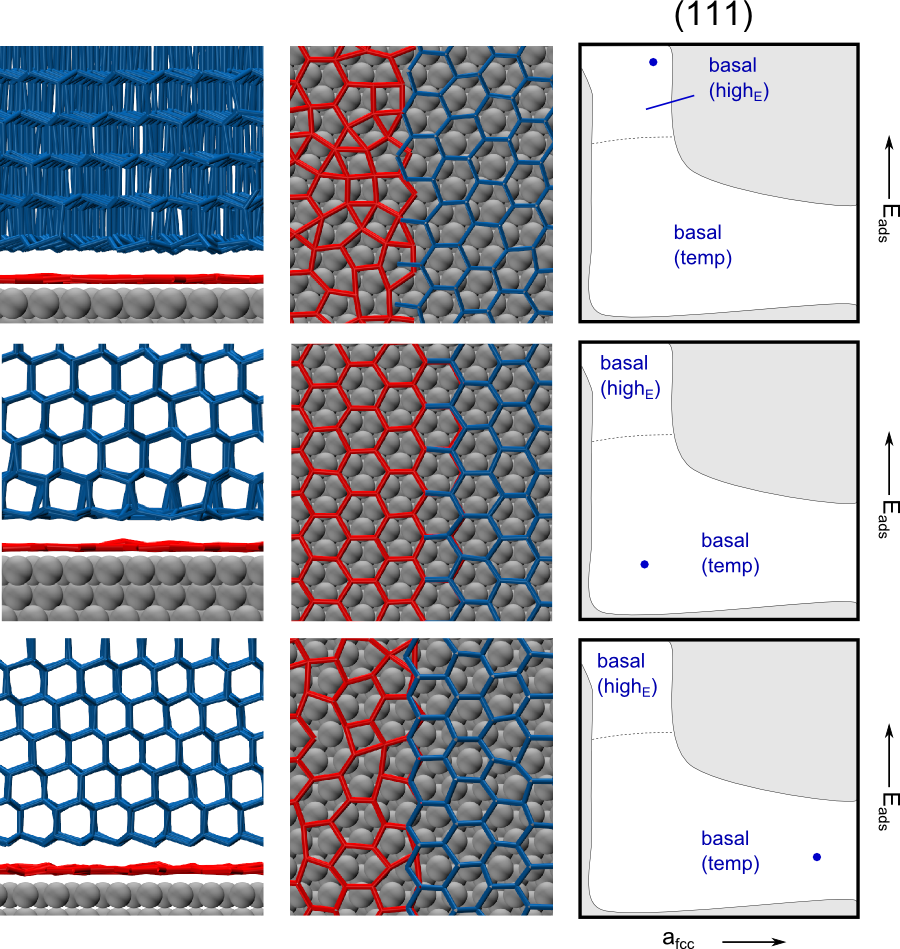}} 
\caption{Classified regions for the 111 surface. Snapshots are taken from regions indicated by the blue dot.} 
\label{FIG_SI_REGIONS_111} 
\end{figure}

\begin{figure}[H]
\centerline{\includegraphics[width=10cm]{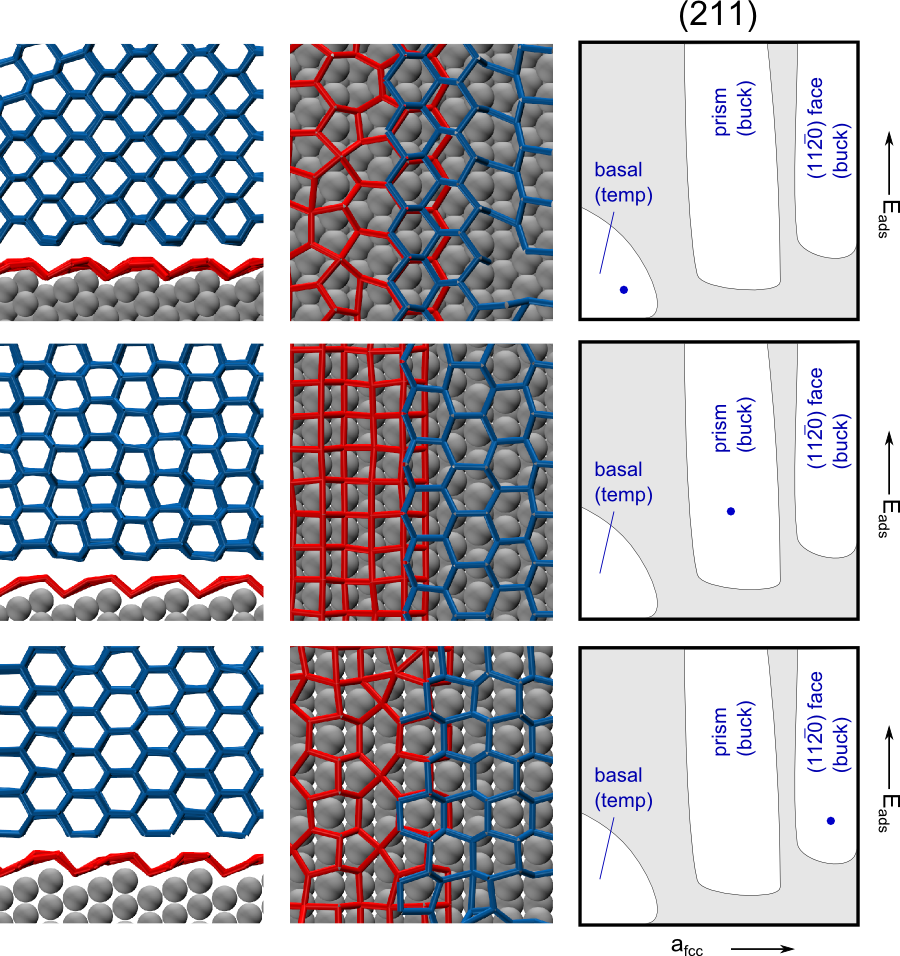}} 
\caption{Classified regions for the 211 surface. Snapshots are taken from regions indicated by the blue dot.} 
\label{FIG_SI_REGIONS_211} 
\end{figure}

\begin{figure}[H]
\centerline{\includegraphics[width=10cm]{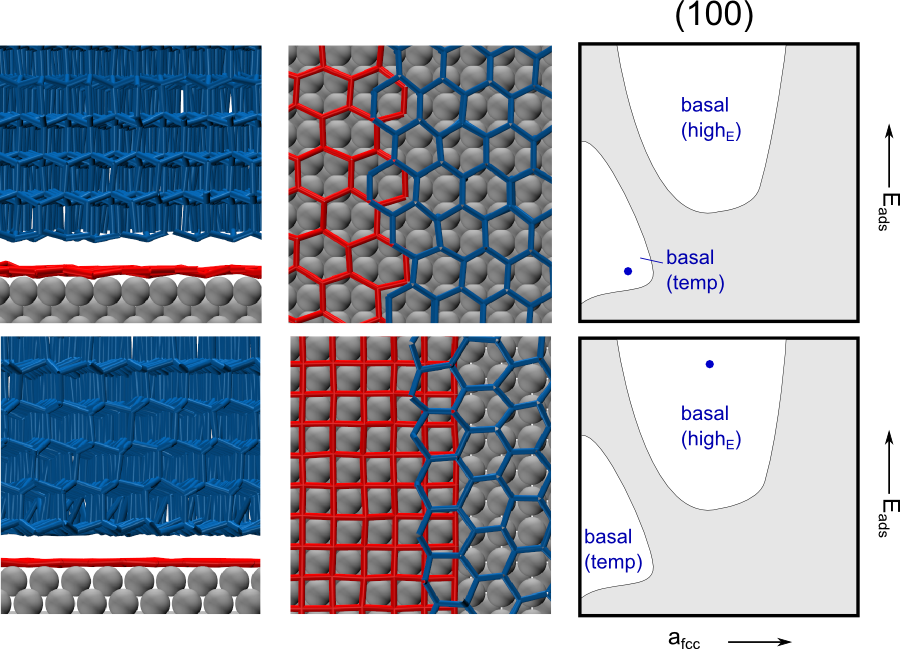}} 
\caption{Classified regions for the 100 surface. Snapshots are taken from regions indicated by the blue dot.} 
\label{FIG_SI_REGIONS_100} 
\end{figure}

\begin{figure}[H]
\centerline{\includegraphics[width=10cm]{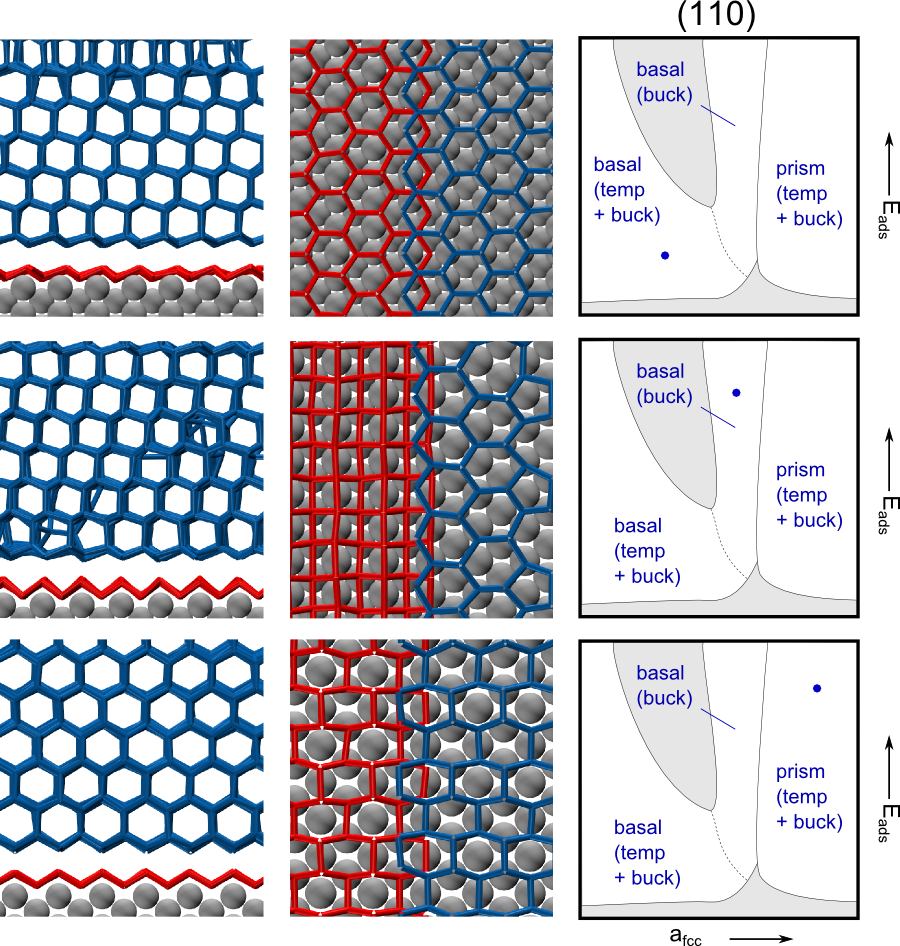}} 
\caption{Classified regions for the 110 surface. We note that the side view of the last region does not show the typical side perspective of the prism face. However, it results from a $90^\circ$ rotation around the z axis of the typical view, as seen e.g. in the prism region of the (211) surface. Snapshots are taken from regions indicated by the blue dot.} 
\label{FIG_SI_REGIONS_110} 
\end{figure}

\clearpage
\newpage
\subsection{Distribution of Pre-Critical Nuclei}
\begin{figure}[h]
\centerline{\includegraphics[width=9cm]{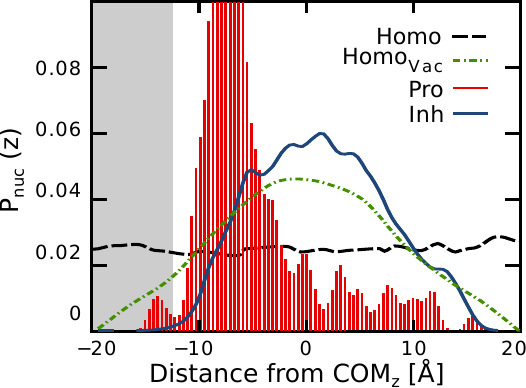}} 
\caption{Probability density distribution $P_\mathrm{nuc}(z)$ of the z-coordinate of center of mass (COM) of pre-critical ice-like clusters. The x-axis refers to the distance from the COM of the mW water slab. The gray shaded region highlights the extent of the 1st and 2nd water overlayer on top of the LJ surface. The legend refers to a bulk model of 4000 mW molecules (Homo), the same as a free-standing slab (Homo$_\mathrm{VAC}$) and scenarios (Inh and Pro) in which we observe inhibition/promotion of $J$ on the (100) surface ($a_\mathrm{fcc}=3.90$~\r{A} for both, $E_\mathrm{ads}= 3.21$ and $5.30$~kcal/mol respectively). All data was collected at 205~K.}
\label{FIG_9_INHIB_2ND_LAYER} 
\end{figure}
\begin{figure}[h]
\centerline{\includegraphics[width=10cm]{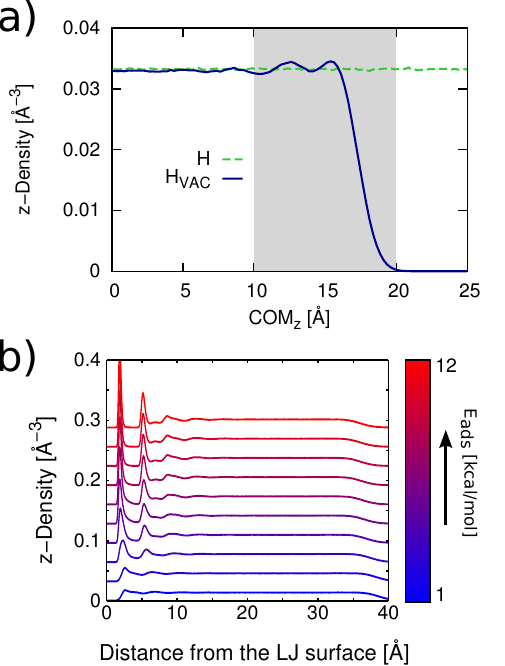}} 
\caption{a) z-Density profile of a homogeneous mW model (4000 molecules) $H$ and the same model as a free standing slab ($H_\mathrm{VAC}$), featuring two water-vacuum interfaces. The x-axis refers to the distance from the center of mass of the mW slab along the z coordinate (only one side of the slab is shown). The shaded region highlights the fraction of the system affected by the presence of the water-vacuum interface because of density oscillations. b) z-Density profile of mW water on top of the (100) surface ($a_\mathrm{fcc}=3.9$~\r{A}, obtained at 290~K) as a function of $E_\mathrm{ads}$.}
\label{FIG_SI_HVAC} 
\end{figure}

The distribution of pre-critical nuclei in Figure~\ref{FIG_9_INHIB_2ND_LAYER} is helpful for discussing what is the only common feature found for all of the four surfaces: inhibition of the nucleation rate for the smallest value of $E_\mathrm{ads}$. For this interaction strength the molecules essentially face a hard wall which in turn could even hinder nucleation compared to the homogeneous case~\cite{reinhardt_effects_2014,haji-akbari_suppression_2014,zhang_freezing_2014,zhang_wall-induced_2014}. To understand this we must first mention, what happens for the homogeneous bulk case (which we term Homo) and the case of a free standing water slab (called Homo$_\mathrm{VAC}$) with two water-vacuum interfaces. The nucleation rate of Homo$_\mathrm{VAC}$ will be lower than in the bulk case (Homo), which can be roughly rationalized in terms of a smaller volume available for the nuclei to appear than in the bulk case. This effect is visible in Figure~\ref{FIG_9_INHIB_2ND_LAYER}, where the distribution for Homo corresponds to a constant line, while the probability for Homo$_\mathrm{VAC}$ is decreased towards the interface. In fact, the nucleation rate constant for Homo$_\mathrm{VAC}$ computed by excluding the volume of the system affected by the presence of the water-vacuum surface (which can be estimated by looking at the density profile along the z-coordinate, see SI Figure~\ref{FIG_SI_HVAC}a) is basically the same as obtained for Homo. In the case of our models, the presence of the LJ surface could introduce significant density perturbations in the water film for all values of $E_\mathrm{ads}$ (see SI Figure~\ref{FIG_SI_HVAC}b). As a result, when no efficient template can be provided by the surface, pre-critical nuclei tend to strictly avoid the neighborhood of the LJ surface as well, as reported in Figure~\ref{FIG_9_INHIB_2ND_LAYER} (Inh). This effect could be even stronger than the inhibition coming from the water-vacuum interface, and as a result the effective volume available for the nuclei to appear is even less than in the Homo$_\mathrm{VAC}$ case, thus causing a net inhibiting effect due to the presence of the surface. It is worth noting that while the promotion of the nucleation rate observed for many ($a_\mathrm{fcc}$,$E_\mathrm{ads}$) points can be rather strong, the inhibition effect is usually much weaker, as it basically accounts for the removal of the portion of the system affected by the presence of the LJ surface. This kind of inhibition is therefore unlikely to be visible in simulations or experiments where the ratio of water volume to contact area is much higher than in our case.

\subsection{Notes on the Water Model}
\begin{figure}[h]
\centerline{\includegraphics[width=14cm]{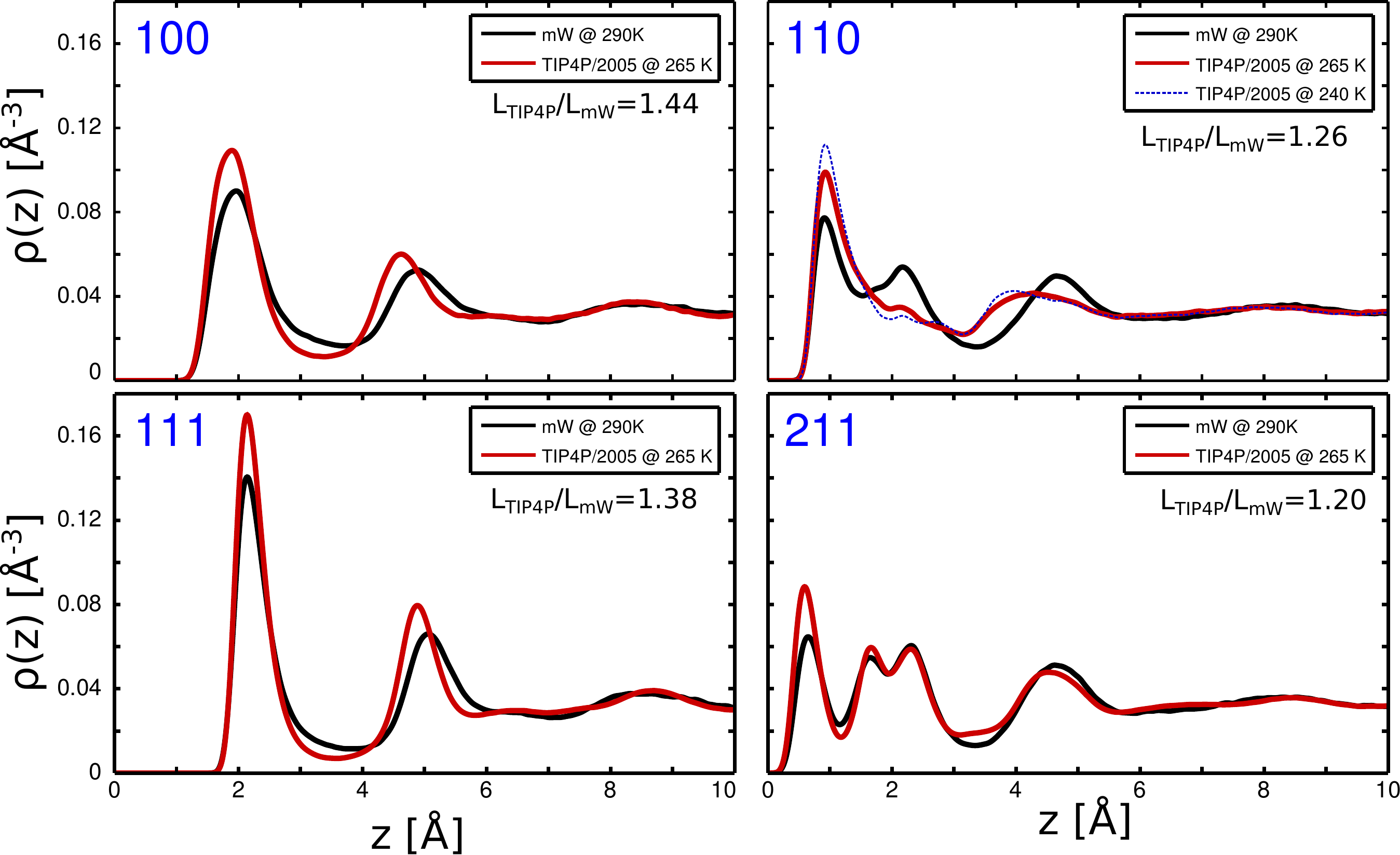}} 
\caption{Comparison of the density perpendicular to the surface for the mW and TIP4P/2005 models of water. Results are based on at least 75~ns long equilibration trajectories approximately 15~K above the melting point of the corresponding model. All graphs were computed for $a_\mathrm{fcc} \approx 3.9$~\r{A} and $E_\mathrm{ads} \approx 3.2$~kcal/mol. The temperature was chosen to be approximately 15~K above the melting point of the respective water model.} 
\label{FIG_SI_DENSITIES} 
\end{figure}

Firstly, we checked that the water densities observed (e.g. depicted in Figure 6) are not an artifact of the coarse grained water model by comparing them to the results of TIP4P/2005. The test was done for one ($a_\mathrm{fcc}$,$E_\mathrm{ads}$) point on each surface. The water-surface interaction was kept identical to the mW case, i.e. the surface atoms interact with the TIP4P/2005 oxygens through the same LJ potential. The resulting densities (see Figure~\ref{FIG_SI_DENSITIES}) are very similar. While the the latter shows slightly stronger peaks and layering, the peak positions agree with the mW values. We conclude that the structuring exhibited by the mW model is nearly equivalent to the one of TIP4P/2005. Our results concerning the buckling and the structuring perpendicular to the surface should therefore be applicable to all-atom models of water. It appears that potential differences for overlayer patterns obtained from simulations are artifacts of the very different time scales on which both models evolve rather than actual structural differences. 

\begin{figure}[h] 
\centerline{\includegraphics[width=8.5cm]{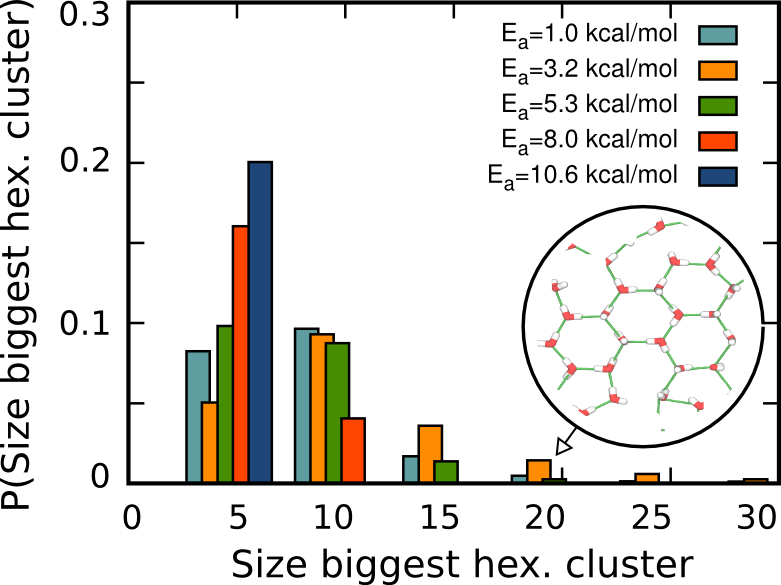}}
\caption{Probability density distribution of the size of the biggest hexagonal patch of TIP4P/2005 water molecules within the first overlayer for different values of $E_\mathrm{ads}$ and $a_\mathrm{fcc}$=3.90~\r{A} on the (111) surface. Results were taken from 20~ns trajectories sampled with a 2~fs timestep. The inset depicts a fairly large cluster of about 20 molecules.} 
\label{FIG_TIP_CLUSTERS} 
\end{figure}

Secondly, we performed nucleation simulations with TIP4P/2005 water on the (111) surface for $a_\mathrm{fcc}$=3.90~\r{A} and different values of $E_\mathrm{ads}$. Contrary to the mW case, using the all-atom model we did not observe the formation of a complete hexagonal overlayer within $100$~ns. This might not seem a surprise, specifically because of the lack of a hydrogen bond network. This deficiency results in a much faster dynamics (we have estimated a mismatch in the self-diffusion coefficient of about three orders of magnitude at the supercooling considered here) of the water molecules with respect to both experiments and basically any full atomistic water model~\cite{molinero_water_2009,orsi_comparative_2014}. Besides, the mW model potential energy surface is much smoother than one in which hydrogen bonds would be taken into account. However, the number and the size of hexagonal patches within the first overlayer is consistent with what we have observed in the case of the mW model. In Figure~\ref{FIG_TIP_CLUSTERS} we report the probability density function of the size of the biggest hexagonal patch of TIP4P/2005 water molecules on top of the (111) surface.  The tails of the distributions, corresponding to sizable hexagonal patches can only be observed for $E_\mathrm{ads}=3.2$~kcal/mol, which is exactly the value for which we observe the fastest formation of the hexagonal overlayer in the case of mW. To gauge interaction energies for different water models the heat of vaporization is often used. The latter for both models is nearly the same~\cite{molinero_water_2009,abascal_general_2005} which means we can compare the adsorption energies directly. The trend holds for different supercooling as well, and confirms that while - especially heterogeneous - nucleation processes modeled by mW water are for certain nonphysically fast, the model could still capture part of the underlying physics of the problem.

\clearpage
\newpage
\subsection{Higher Temperatures}
\begin{figure}[h]
\centerline{\includegraphics[width=8.5cm]{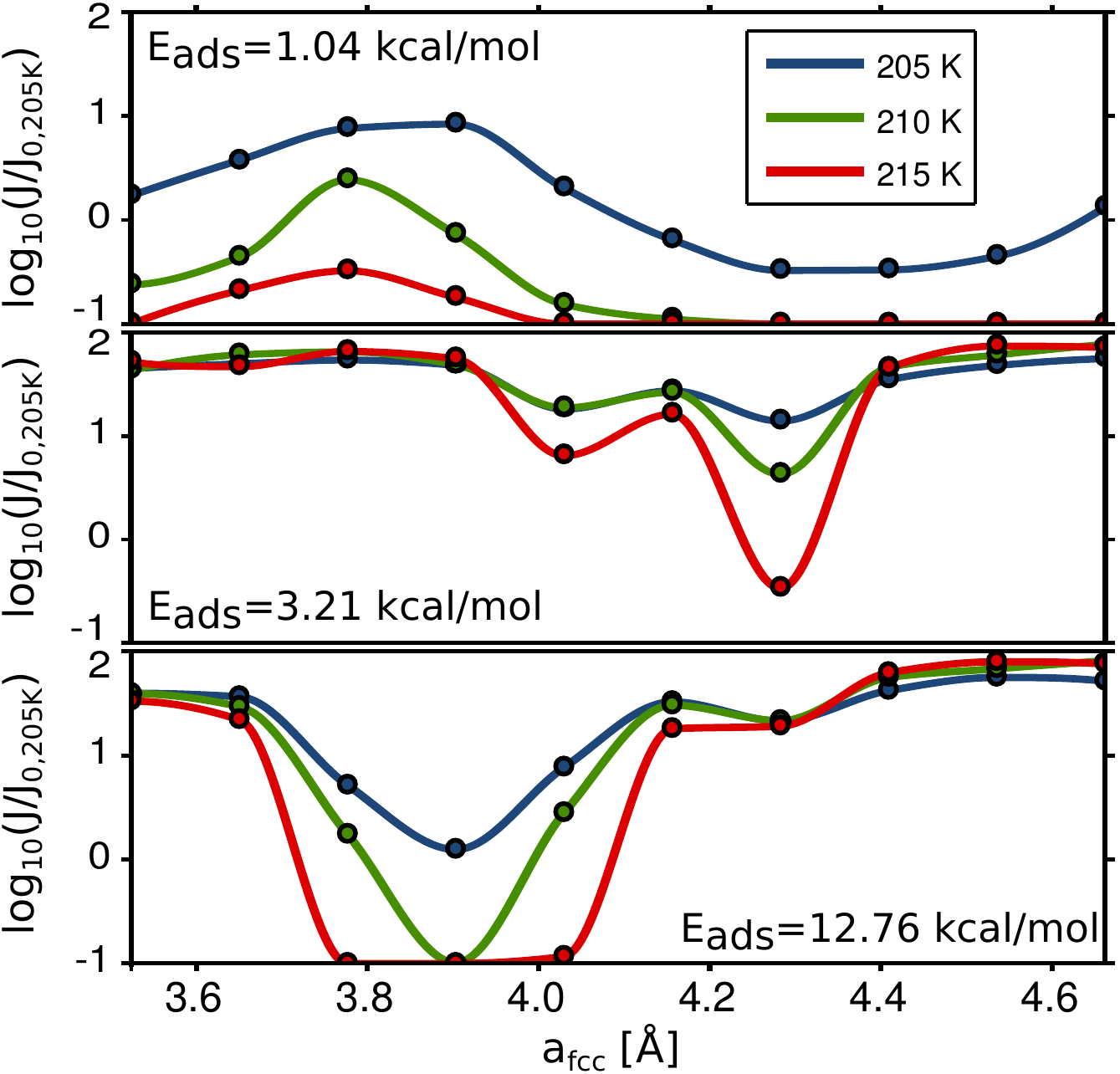}} 
\caption{Temperature dependence of nucleation rates (circles) and spline interpolation (colored lines) for 3 adsorption energies on the (110) surface. All values were normalized by the homogeneous nucleation rate $J_0$ at 205~K.} 
\label{FIG_6_DIFFERENT_TEMPERATURES} 
\end{figure}

Our simulations have been performed in the deeply supercooled regime. It is beyond reach to do such an extensive set of simulations as performed here at significantly higher temperature. Nonetheless it is interesting to understand how the phenomena observed might depend on temperature. To estimate the effect of the strong supercooling on the results and especially the proposed mechanisms we performed calculations at higher temperatures for three adsorption energies on the (110) surface (depicted in Figure~\ref{FIG_6_DIFFERENT_TEMPERATURES}). Since the computational cost significantly increases, the nucleation rates become less accessible with the brute force approach and we therefore limit this trial to only a few ($a_\mathrm{fcc}$,$E_\mathrm{ads}$) points. As previously mentioned, the values at the top and bottom range of the nucleation rate should be seen as a lower/upper bound to the actual nucleation rate. Conclusively, a missing temperature dependence of these points indicates a nucleation rate out of the limit that can be resolved with simulations of 500~ns length, rather than one that is constant with temperature. We find that the trends seen at 205~K are stable against the temperature increase and only in regions where no specific mechanism has been attributed the rates heavily decline. In fact the increased temperature can help to identify the values of $a_{fcc}$ inducing a certain mechanism in a more precise way because the gaps between the enhanced regions increase. Furthermore, the structures of the adsorption layers did not show any noticeable change so that we can assume that our conclusions are valid also for higher temperatures. 


\end{document}

%% file: paper.bbl
%

%% file: main.ARXIV.bbl
\begin{thebibliography}{137}
\bibitem [{not({\natexlab{f}})}]{note_largerDifference}%
  \BibitemOpen
  \href@noop {} {}\bibinfo {note} {{T}he actual difference in nucleation rates
  for Au(111) and Au(100) could be even larger, since the highest nucleation
  rates computed in our study are only lower bounds. {T}herefore a significant
  difference between these two substrates should also be detectable if one
  measures freezing temperatures.}\BibitemShut {Stop}%
\end{thebibliography}
